\documentstyle[amsmath,amssymb,epsfig,12pt]{article}

\begin{document}
\setlength{\parskip}{0.45cm}
\setlength{\baselineskip}{0.75cm}
%XXXXXXXXXXXXXXXXXXXXXXXXXXXXXXXXXXXXXX
%
%SETTINGS FOR PREPRINT-SPACED VERSION
%setlength{\parskip}{0.45cm}
%setlength{\baselineskip}{0.75cm}
%
% SETTINGS FOR DOUBLE - SPACED VERSION
%\setlength{\parskip}{0.65cm}
%\setlength{\baselineskip}{0.95cm}
%
%XXXXXXXXXXXXXXXXXXXXXXXXXXXXXXXXXXXXXX
\begin{titlepage}
\setlength{\parskip}{0.25cm}
\setlength{\baselineskip}{0.25cm}
\begin{flushright}
DO-TH 06/05\\
\vspace{0.2cm}
%astro--ph/0303252\\
%\vspace{0.2cm}
July 2006
\end{flushright}
\vspace{1.0cm}
\begin{center}
\Large
{\bf On the transport equations of cosmic neutrinos}
\\\Large{\bf passing through Earth and secondary $\nu_\mu$ fluxes}
\vspace{1.5cm}

\large
S.~Rakshit and E.\ Reya\\
\vspace{1.0cm}

\normalsize
{\it Universit\"{a}t Dortmund, Institut f\"{u}r Physik,}\\
{\it D-44221 Dortmund, Germany} \\
\vspace{0.5cm}

\vspace{1.5cm}
\end{center}

\begin{abstract}
\noindent The convergence of the iterative solutions of the transport 
equations of cosmic muon and tau neutrinos propagating through Earth is
studied and analyzed.  For achieving a fast convergence of the iterative
solutions of the coupled transport equations of 
$\nu_\tau$, $\bar{\nu}_\tau$
and the associated $\tau^{\pm}$ fluxes, a new semi--analytic input
algorithm is presented where the peculiar $\tau$--decay contributions are
implemented already in the initial zeroth order input.  Furthermore,
the common single
transport equation for muon neutrinos is generalized by taking into account
the contributions of secondary $\nu_\mu$ and $\bar{\nu}_\mu$ fluxes due to  
the prompt 
$\tau$--decay $\tau\to \nu_\mu$ initiated by the associated tau flux.
Differential and total nadir angle
integrated upward--going $\mu^-+\mu^+$ event rates are presented for 
underground neutrino telescopes and compared with the muon rates initiated
by the primary $\nu_\mu$, $\nu_\tau$ and $\tau$ fluxes.
\end{abstract}
\end{titlepage}

%MAIN PART

%\renewcommand{\theequation}{\arabic{section}.\arabic{equation}}
\section{Introduction}
Upward--going cosmic neutrinos with energies below $10^8$ GeV play a 
decisive role for underground neutrino telescopes, since the atmospheric
background can be more effectively controlled, in contrast to downward--going
cosmic neutrinos.  While traversing through the Earth, upward--going muon 
(anti)neutrinos undergo attenuation (absorption) due to weak charged current
(CC) and neutral current (NC) interactions as well as regeneration
\cite{ref1,ref2} due to NC interactions.  The latter shift the energy of
the neutrinos, rather than absorbing them, to lower energies and populate
the lower energy part of the initial cosmic neutrino flux spectra, thus
adding to the naive non--regenerated $\mu^- +\mu^+$ event rates at the 
detector.  Such propagation effects of muon (and electron) neutrinos through
Earth are described by a single transport (integro--differential) equation
which can be rather easily solved iteratively \cite{ref1,ref2,ref3,ref4,ref5}.

On the other hand tau (anti)neutrinos are not absorbed, but degraded in
energy, in the Earth as long as the interaction length of the produced
tau leptons is larger than their decay length (which holds for energies up
to about $10^9$ GeV).  Because of these latter (semi)leptonic decays
$\tau\to\nu_\tau X$, the Earth will not become opaque to $\nu_\tau$
\cite{ref6} since the $\tau^-$ produced in CC interactions decays back
to $\nu_\tau$.  This `regeneration chain' $\nu_\tau\to\tau\to\nu_\tau\to\ldots$
continues until the $\nu_\tau$ and $\bar{\nu}_\tau$, as well as the 
$\tau^{\pm}$ leptons, reach the detector on the opposite side of the Earth.
Thus the propagation of high--energy tau neutrinos through the Earth is
very different from muon and electron neutrinos, and we have now to deal
with coupled transport equations for the $\stackrel{(-)}{\nu}_\tau$ and
$\tau^{\pm}$ fluxes \cite{ref4,ref7,ref8,ref9,ref10,ref11,ref12,ref13}.
Obtaining stable iterative solutions of these coupled integro--differential
equations is far more involved as compared to the single transport equation
for muon neutrinos.  It is one of our main objectives to discuss the general
qualitative and quantitative structure of these solutions and to present
an efficient input algorithm which allows for a rather fast convergence of
the iterative procedure.  This applies to all present model cosmic neutrino
fluxes.  Moreover the $\tau^- +\tau^+$ flux, generated by the initial cosmic
$\nu_\tau+\bar{\nu}_\tau$ flux while traversing the Earth, gives rise to a
secondary $\bar{\nu}_\mu+\nu_\mu$ flux \cite{ref14} via $\tau\to\nu_\mu$ 
due to the prompt $\tau$--decays like $\tau^-\to\nu_\tau \mu^-\bar{\nu}_\mu$.
This adds considerable contributions to the primary cosmic 
$\nu_\mu+\bar{\nu}_\mu$ flux and may increase the $\mu^- +\mu^+$ rates at
the detector site sizeably \cite{ref9,ref12}, depending on the cosmic flux
and nadir angle considered.  Such effects require an extension of the simple
single transport equation for  $\stackrel{(-)}{\nu}_\mu$ and the inclusion
of the appropriate prompt decay term reduces the convergence of the iterative
procedure considerably. 

The simple single transport equation for  $\stackrel{(-)}{\nu}_\mu$ will be
discussed for completeness in Sec.~2.  Although frequently used, the 
excellent convergence of its iterative solutions has not been explicitly
demonstrated thus far for more realistic cosmic neutrino fluxes, apart from
some specific steep toy model neutrino fluxes \cite{ref2}.  In Sec.~3
we turn to the iterative solutions of the far more complicated coupled
transport equations for  $\stackrel{(-)}{\nu}_\tau$ and their associated
$\tau^{\pm}$ fluxes.  A new semi--analytic input algorithm is presented 
which allows for a fast convergence of the iterative solutions. 
The implications for the upward--going $\mu^- +\mu^+$ event rates for 
underground neutrino detectors for some relevant cosmic neutrino fluxes will
be briefly outlined as well. The solutions of the generalized single 
transport equation for muon neutrinos, by taking into account the contributions
of the secondary $\nu_\mu +\bar{\nu}_\mu$ flux from prompt $\tau^{\pm}$
decays based on the calculated associated
$\tau^{\pm}$ fluxes, are discussed in Sec.~4.  Their implications for the 
expected $\mu^- +\mu^+$ event rates, as initiated by various relevant cosmic
neutrino model fluxes, are presented as well.  Finally, our
conclusions are summarized in Sec.~5.

\section{The transport equation of muon neutrinos}

Disregarding possible contributions from other neutrino flavors for the 
time being, the transport equation for upward--going cosmic muon 
(anti)neutrinos $\stackrel{(-)}{\nu}_\mu$ passing through Earth can be 
written as \cite{ref1,ref2,ref3,ref4,ref5}
%Eq.(1)
\begin{equation}
\frac{\partial F_{\nu_\mu}(E,X)}{\partial X} = 
  -\frac{F_{\nu_\mu}(E,X)}{\lambda_\nu(E)} +\frac{1}{\lambda_\nu(E)}
    \int_0^1 \frac{dy}{1-y}\, K_\nu^{\rm NC}(E,y)\, F_{\nu_\mu}(E_y,X)
\end{equation}
where $F_{\nu_\mu}\equiv d\Phi_{\nu_\mu}/dE$ is the differential cosmic 
neutrino flux and $E_y=E/(1-y)$.  The column depth $X=X(\theta)$, being the
thickness of matter traversed by the upgoing leptons, depends on the nadir
angle of the incident neutrino beam ($\theta=0^{\rm o}$ corresponds to a 
beam traversing the diameter of the Earth); it is obtained from integrating
the density $\rho(r)$ of the Earth along the neutrino beam path $L'$ at a
given $\theta$, $X(\theta)=\int_0^L\rho (L')dL'$ with 
$L= 2 R_{\oplus}\cos\theta$, $R_{\oplus}\simeq 6371$ km, denoting the 
position of the underground detector, and $X(\theta)$ can be found, 
for example, in Fig.~15 of \cite{ref15} in units of g/cm$^2$ = cm we.
Furthermore $\lambda_\nu^{-1} = N_A \sigma_{\nu N}^{\rm tot}$, 
$N_A = 6.022 \times 10^{23} g^{-1}$, is the inverse neutrino interaction
length where $\sigma_{\nu N}^{\rm tot} = \sigma_{\nu N}^{\rm CC} 
+\sigma_{\nu N}^{\rm NC}$ and
 %Eq.(2)
\begin{equation}
K_\nu^{\rm NC}(E,y) = \frac{1}{\sigma_{\nu N}^{\rm tot}(E)} \,\,
     \frac{d\sigma_{\nu N}^{\rm NC}(E_y,y)}{dy}\, .
\end{equation}
The various CC and NC $\stackrel{(-)}{\nu}\!\!N$ cross sections are calculated as
in \cite{ref5,ref13}, with the relevant details to be found in \cite{ref16},
utilizing the QCD inspired dynamical small--$x$ predictions for parton 
distributions according to the radiative parton model \cite{ref17}.  Notice
that conventionally fitted parton distributions at the relevant weak scale
$Q^2=M_W^2$ would require additional ad hoc assumptions (see, e.g., 
\cite{ref15,ref18}) for the necessary extrapolations into the yet unmeasured
small Bjorken--$x$ region $x<10^{-3}$ ($x\simeq M_W^2/2m_N E$).  The first
term in (1) describes the attenuation (absorption) of neutrinos when
penetrating through the Earth, and the second one their regeneration 
consisting of the degrading shift in their energy.  For definiteness all
formulae are given for an incoming neutrino beam, but similar expressions
hold of course for antineutrinos.

Equation (1) can be efficiently solved by the ansatz \cite{ref2}
%Eq.(3)
\begin{equation}
F_{\nu_\mu}(E,X) 
       = F_{\nu_\mu}^0(E)\exp \left[-\frac{X}{\Lambda_{\nu_\mu}(E,X)}\right]
\end{equation}
with an effective absorption (interaction) length
%Eq.(4)
\begin{equation}
\Lambda_{\nu_\mu}(E,X) = \frac{\lambda_\nu(E)}{1-Z_{\nu_\mu}(E,X)}
\end{equation}
and where $F_{\nu_\mu}^0(E)\equiv F_{\nu_\mu}(E,X=0)$ denotes the initial 
cosmic neutrino flux which reaches the Earth's surface.  Depending on the 
assumed cosmic neutrino flux, the $Z_\nu$--factor can take any non--negative
values.  Its physics interpretation and the consequences for the shadowing
factor 
$S\equiv\exp\left[ -X/\Lambda_\nu\right]$ in (3) are immediate: $Z_\nu<1$
(the only case considered in \cite{ref2} relevant for steeper, i.e., soft
model fluxes) implies $\Lambda_\nu>\lambda_\nu>0$ thus $S<1$, i.e.\ the
neutrino flux will be further attenuated since absorption plays the dominant
role; for $Z_\nu =1$, $\Lambda_\nu=\infty$, i.e.\ $S=1$ which means that 
regeneration and absorption compensate each other; finally $Z_\nu>1$ implies
$\Lambda_\nu<0$ and $S>1$, and consequently the NC regeneration in (1) can
even cause an enhancement of the neutrino spectrum with respect to the 
initial flux $F_{\nu_\mu}^0(E)$ for certain energies and depths $X$.  
Inserting (3) into (1) yields
%Eq.(5)
\begin{equation}
Z_{\nu_\mu}(E,X) =\frac{1}{X}\int_0^X dX' \int_0^1 dy\,  K_\nu^{\rm NC}(E,y)\,
   \eta_\nu\, (E,y)\, e^{-X'D_{\nu_\mu}(E,E_y,X')}
\end{equation}
with $\eta_\nu(E,y)=F_{\nu_\mu}^0(E_y)/(1-y)F_{\nu_\mu}^0(E)$ and $D_{\nu_\mu}
(E,E_y,X') = \Lambda_{\nu_\mu}^{-1}(E_y,X')-\Lambda_{\nu_\mu}^{-1}(E,X')$.
Using an iteration algorithm to solve for $Z_{\nu_\mu}(E,X)$, one can
formally rewrite the solution of (5) after the n--th iteration as
%Eq.(6)
\begin{equation}
Z_{\nu_\mu}^{(n+1)}(E,X) = \frac{1}{X} \int_0^X dX' \int_0^1 dy\, 
  K_\nu^{\rm NC}(E,y)\, \eta_\nu(E,y)\, e^{-X'D_{\nu_\mu}^{(n)}(E,E_y,X')}
\end{equation}
where
%Eq.(7)
\begin{equation}
D_{\nu_\mu}^{(n)}(E,E_y,X') 
   = \frac{ 1-Z_{\nu_\mu}^{(n)}(E_y,X')}{\lambda_\nu(E_y)}
- \frac{1-Z_{\nu_\mu}^{(n)}(E,X')}{\lambda_\nu(E)} \,\, .
\end{equation}
The reason why this iteration is expected to converge very fast is as
follows:  the kernel $K_\nu^{\rm NC}$ peaks very strongly \cite{ref2,ref19}
at $y=0$ and $y=1$, with the contribution at $y\simeq 1$ being, however,
exponentially suppressed in (6); thus the main contribution to the 
integral over $y$ in (6) comes from the region around $y\simeq 0$ where
$D_{\nu_\mu}(E, E_y, X')\to 0$ as $y\to 0$.  Therefore the iteration 
should be robust with respect to choosing the $n=0$ approximation 
\cite{ref2}.  The most simple input choice is $Z_{\nu_\mu}^{(0)}(E,X')=0$
in (7).  For this case the analytic $X'$--integration in (6) yields
%Eq.(8)
\begin{equation}
Z_{\nu_\mu}^{(1)}(E,X)=\int_0^1 dy K_\nu^{\rm NC}(E,y)\, \eta_\nu (E,y)\,
  \frac{1-e^{-XD_\nu(E,E_y)}}{XD_\nu(E,E_y)}
\end{equation}
with
%Eq.(9)
\begin{equation}
D_\nu(E,E_y) \equiv D_{\nu_\mu}^{(0)}(E,E_y,X') = 
   \frac{1}{\lambda_\nu(E_y)} - \frac{1}{\lambda_\nu(E)}\,\, .
\end{equation}
With the $n=1$ solution in (8) at hand, it is now straightforward to
obtain iterations in higher orders, for example, for $n=2$ by inserting
(8) into (7) gives $Z_{\nu_\mu}^{(2)}$ in (6).

Representative cosmic neutrino fluxes of some hypothesized sources are 
displayed in Fig.~1 which we shall partly use for all our subsequent
calculations.  Recent diffuse neutrino flux upper limits of AMANDA 
\cite{ref20,ref21} are shown by the bars with arrows -- the latter 
indicate the still allowed region.  Although the huge flux from active
galactic nuclei of Stecker and Salamon (AGN--SS) \cite{ref22} has been
already excluded, we shall use it merely as a theoretical playground
due to its unique spectrum at lower energies where 
$F_{\nu_\mu}^0(E)\sim$ const.\ for 
$E$ \raisebox{-0.1cm}{$\stackrel{<}{\sim}$} $10^5$ GeV.
On the other hand the AGN--M95 flux \cite{ref23} is still compatible
(although slightly in conflict) with the AMANDA upper bound, as are
the gamma ray burst (GRB--WB) \cite{ref24} and topological defect
(TD--SLBY) \cite{ref25} fluxes.  These latter three fluxes will be used
for our  `realistic' model calculations.  The TD--SLSC \cite{ref26} and
$Z$--burst \cite{ref27} fluxes are shown just for illustration since
they are too minute for being tested with upward--going event rates
\cite{ref5}.  Note that the initial cosmic (anti)neutrino fluxes
$F_{\nu,\bar{\nu}}^0(E)$ in (3) which reach the Earth's surface are given 
by $F_{\nu_\mu}^0 = F_{\bar{\nu}_\mu}^0 = F_{\nu_\tau}^0 = 
F_{\bar{\nu}_\tau}^0 = \frac{1}{4}d\Phi/dE$ with $\Phi$ being the 
cosmic $\nu_\mu+\bar{\nu}_\mu$ flux at the production site in Fig.~1.  

For a better comparison of our quantitative results with the ones 
obtained in the literature, we also employ two generic initial fluxes
incident on the surface of the Earth at a nadir angle $\theta =0^{\rm o}$
of the form \cite{ref4,ref7}
%Eqs.(10)+(11)
\begin{eqnarray}
F_{\nu_\mu+\bar{\nu}_\mu}^0(E) & = & N_1 E^{-1}(1+E/E_0)^{-2},\quad
   E_0 = 10^8\,\,{\rm GeV}\\
F_{\nu_\mu+\bar{\nu}_\mu}^0(E) & = & N_2 E^{-2}
\end{eqnarray}
with adjustable normalization factors $N_i$, for example, $N_1=\frac{1}{2}
\times 10^{-13}/$(cm$^2$  sr s) and $N_2=\frac{1}{2}\times 10^{-7}$
GeV/(cm$^2$ sr s). Notice that the generic $E^{-1}$ energy dependence
is representative for the TD and $Z$--burst fluxes in Fig.~1 for 
$E$ \raisebox{-0.1cm}{$\stackrel{<}{\sim}$} $10^7$ GeV; and also for the
GRB--WB flux for $E$
\raisebox{-0.1cm}{$\stackrel{<}{\sim}$} $10^5$ GeV. Furthermore
the latter GRB--WB flux behaves like $E^{-2}$ in (11) for 
$10^5 < E$ \raisebox{-0.1cm}{$\stackrel{<}{\sim}$} $10^7$ GeV,
where such a power spectrum with index $-2$ is typical for shock
acceleration (see, e.g., \cite{ref21}).

Our results for $Z_{\nu_\mu}^{(1)}$ and $Z_{\nu_\mu}^{(2)}$ are shown in
Figs.~2 and 3 for two typical values of the nadir angle, 
$\theta =0^{\rm o}$ ($X=1.1\times 10^{10}$ cm we) and $\theta =
50^{\rm o}$ ($X=3.6\times 10^{9}$ cm we).  The iteration converges very
fast since in general the maximum difference between $Z_{\nu_\mu}^{(1)}$
and $Z_{\nu_\mu}^{(2)}$ is less than about 5\%, $|Z_{\nu_\mu}^{(2)}/
Z_{\nu_\mu}^{(1)}-1|<0.05$, and moreover $|Z_{\nu_\mu}^{(3)}/
Z_{\nu_\mu}^{(2)}-1|<0.005$.  Thus the first $n=1$ iteration is already
sufficiently stable and suffices for {\em all} cosmic neutrino fluxes
considered at present \cite{ref19}.  Notice that for larger $\theta$
(smaller $X$) the difference between $Z_{\nu_\mu}^{(2)}$ and 
$Z_{\nu_\mu}^{(1)}$ decreases and therefore the stability increases.
The results for $Z_{\bar{\nu}_\mu}$ are similar but $Z_{\bar{\nu}_\mu}
>Z_{\nu_\mu}$ for 
$E$ \raisebox{-0.1cm}{$\stackrel{<}{\sim}$} $10^6$ GeV where 
$\lambda_{\bar{\nu}}>\lambda_{\nu}$.  The resulting $\nu_{\mu}$ and
$\bar{\nu}_{\mu}$ fluxes follow from (3) and can be found in 
\cite{ref4,ref5,ref7}.

\section{The transport equations of tau neutrinos and taus}

Apart from the absorption (attenuation) due to $\sigma_{\nu N}^{\rm tot}$
and regeneration due to $\sigma_{\nu N}^{\rm NC}$ in (1), for 
upward--going cosmic tau (anti)neutrinos $\stackrel{(-)}{\nu}_\tau$,
it is important to take into account the regeneration from the 
$\tau^{\pm}$ decays as well as the contributions from the CC tau 
interactions.  The tau neutrino and tau fluxes then satisfy the following
coupled transport equations:
%Eq.(12)+(Eq.(13)
\renewcommand{\arraystretch}{3.0}
\begin{eqnarray}
\frac{\partial F_{\nu_{\tau}}(E,X)}{\partial X} 
 & = & -\frac{F_{\nu_{\tau}}(E,X)}{\lambda_{\nu}(E)} 
    + \frac{1}{\lambda_{\nu}(E)} \int_0^1\frac{dy}{1-y}\,
      K_{\nu}^{\rm NC}(E,y)\, F_{\nu_{\tau}}(E_y,X) \nonumber\\
& & + \int_0^1\frac{dy}{1-y}\, K_{\tau}(E,y)F_{\tau}(E_y,X)\\
\nonumber\\
\frac{\partial F_{\tau}(E,X)}{\partial X} 
 &  = & - \frac{F_{\tau}(E,X)}{\hat{\lambda}(E)} + 
       \frac{\partial\left[\gamma(E)F_{\tau}(E,X)\right]}{\partial E}\nonumber\\
 & &  + \frac{1}{\lambda_{\nu}(E)} \int_0^1 \frac{dy}{1-y}\,
      K_{\nu}^{\rm CC} (E,y)\, F_{\nu_{\tau}}(E_y,X)
\end{eqnarray}
where $F_{\nu_\tau}\equiv d\Phi_{\nu_\tau}/dE$ and 
$F_\tau\equiv d\Phi_\tau/dE$ are the differential energy spectra (fluxes)
of tau (anti)neutrinos and tau leptons and the initial fluxes at the 
surface of the Earth ($X=0$) being given by $F_{\nu_\tau}^0(E) = 
F_{\bar{\nu}_\tau}^0(E) = \frac{1}{4} d\Phi/dE$ with $\Phi$ being the
$\nu_\mu +\bar{\nu}_\mu$ flux at the cosmic production site in Fig.~1.
The cross section kernel $K_{\nu}^{\rm NC}$ is defined in (2) and a
similar expression holds for $K_\nu^{\rm CC}$.  Furthermore
%Eq.(14)
\begin{equation}
K_{\tau}(E,y)= \frac{1}{\lambda_{\tau}(E)}\, K_{\tau}^{\rm CC}(E,y) +
   \frac{1}{\lambda_{\tau}^{\rm dec}(E)}\, K_{\tau}^{\rm dec}(E,y)
\end{equation}
where
\begin{displaymath}
K_{\tau}^{\rm CC}(E,y) = \frac{1}{\sigma_{\tau N}^{\rm tot}(E)}\,\,
   \frac{d\sigma_{\tau N}^{\rm CC}(E_y,y)}{dy},\quad\quad K_{\tau}^{\rm dec}(E,y) 
    =\frac{1}{\Gamma_{\tau}^{\rm tot}(E)}\,\, 
       \frac{d\Gamma_{\tau\to \nu_{\tau}X'}(E_y,y)}{dy}\\
\end{displaymath}
and 
$\lambda_\tau^{-1} = N_A\sigma_{\tau N}^{\rm tot} 
           = N_A(\sigma_{\tau N}^{\rm CC} +\sigma_{\tau N}^{\rm NC})$, 
and 
$\hat{\lambda}^{-1} = (\lambda_{\tau}^{\rm CC})^{-1} +
                                 (\lambda_{\tau}^{\rm dec})^{-1}$
with
$(\lambda_\tau^{\rm CC})^{-1} = N_A\sigma_{\tau N}^{\rm CC}$ in (13).
The decay length of the $\tau^{\pm}$ is 
$\lambda_\tau^{\rm dec}(E,X,\theta)=(E/m_\tau)c\tau_\tau \rho(X,\theta)$
with $m_\tau = 1.777$ GeV, $c\tau_\tau = 87.11$ $\mu$m and $\rho$ denoting
the Earth's density (see, e.g., \cite{ref15}).  Furthermore, since
$1/\Gamma_\tau^{\rm tot}(E) = (E/m_\tau)\tau_\tau$, the $\tau$--decay
distribution in (14) becomes $K_\tau^{\rm dec}(E,y) = (1-y)\, dn(z)/dy$
with $z\equiv E_{\nu_\tau}/E_\tau=E/E_y=1-y$ and \cite{ref7,ref28}
%Eq.(15)
\begin{equation}
\frac{dn(z)}{dy} = \sum_i B_i\left[g_0^i(z)+Pg_1^i(z)\right]
\end{equation}
with the polarization $P=\pm 1$ of the decaying $\tau^{\pm}$.  The
$\tau\to \nu_\tau X'$ branching fractions $B_i$ into the decay channel $i$
and the functions $g_{0,1}^i(z)$ are given in Table I of \cite{ref7}.
The decay channels $i$ considered are $\tau\to\nu_\tau \mu\nu_\mu$,
$\tau\to\nu_\tau\pi$, $\tau\to\nu_\tau\rho$, $\tau\to\nu_\tau a_1$ and
$\tau\to\nu_\tau X$ which have branching fractions of 0.18, 0.11, 0.26,
0.13 and 0.13, respectively.  The lepton energy--loss is treated
continuously \cite{ref29,ref30,ref31} by the term proportional to 
$\gamma(E)$ in (13).  Alternatively, the average energy--loss can be treated 
separately (stochastically) \cite{ref32,ref33}, i.e., not including the term
proportional to $\gamma(E)$ in (13) but using instead $-dE/dX=\gamma(E)
=\alpha+\beta E$.  We shall compare these two approaches for taus and
muons toward the end of this Section.  The most general solution of 
Eqs.~(12) and (13) has been presented in \cite{ref10,ref13}, and in the
context of atmospheric muons in \cite{ref31}.  For the time being,
however, we disregard the $\gamma$-term in (13) since observable 
non--negligible upward--going event rates are obtained only for energies
$E<10^8$ GeV \cite{ref7,ref13} where the energy--loss of the taus can 
be neglected \cite{ref10,ref32,ref33,ref34,ref35}.

In the relevant energy region below $10^8$ GeV, the tau--lepton 
interaction length is much larger than the decay length of the $\tau$
(see, e.g., \cite{ref33} and below), 
$\lambda_\tau(E)\gg \lambda_\tau^{\rm dec}(E)$, i.e., 
$K_\tau \simeq K_\tau^{\rm dec}/\lambda_\tau^{\rm dec}$ in (14) and
$\hat{\lambda}_\tau^{-1} \simeq (\lambda_\tau^{\rm dec})^{-1}$ in (13).
Solving (12) and (13) with a similar ansatz as for muon neutrinos in (3),
we write
%Eq.(16)
\begin{equation}
F_{\nu_\tau}(E,X) = F_{\nu_\tau}^0(E)\exp
     \left[-\frac{X}{\Lambda_{\nu_\tau}(E,X)}\right]
\end{equation}
with an effective interaction (absorption) length
%Eq.(17)
\begin{equation}
\Lambda_{\nu_\tau}(E,X) = \frac{\lambda_\nu(E)}{1-Z(E,X)}
\end{equation}
where $Z=Z_{\nu_\tau}+Z_\tau$.  Inserting (16) into (12) and (13) yields
\cite{ref4,ref13}
%Eq.(18)
\begin{equation}
Z_{\nu_\tau}(E,X) = \frac{1}{X}\int_0^X dX'\int_0^1 dy\, K_\nu^{\rm NC}
  (E,y)\, \eta_\nu(E,y)\, e^{-X'D_{\nu_\tau}(E,E_y,X')}
\end{equation}
with $\eta_\nu$ as in (5) since $F_{\nu_\mu}^0=F_{\nu_\tau}^0$
and $D_{\nu_\tau}(E,E_y,X')=\Lambda_{\nu_\tau}^{-1}(E_y,X')-
\Lambda_{\nu_\tau}^{-1}(E,X')$,
and
%Eq.(19)
\begin{equation}
Z_\tau(E,X) = \frac{\lambda_\nu(E)}{X}\int_0^X dX'\int_0^1 dy\,
  \frac{K_\tau^{\rm dec}(E,y)}{\lambda_\tau^{\rm dec}(E,X')}\, 
     F_\tau(E_y,X')\, \frac{\eta_\nu(E,y)}{F_{\nu_\tau}^0(E_y)}\, 
       e^{X'/\Lambda_{\nu_\tau}(E,X')}
\end{equation}
where the obvious dependence of $\lambda_\tau^{\rm dec}$ on $\theta'=
\theta(X')$ has been suppressed and 
%Eq.(20)
\begin{eqnarray}
F_\tau(E_y,X') & = & \frac{F_{\nu_\tau}^0(E_y)}{\lambda_\nu(E_y)} 
  \int_0^{X'} dX'' \int_0^1 dy'\, K_\nu^{\rm CC} (E_y,y')\, 
    \eta_\nu(E_y,y')\, e^{-X''/\Lambda_{\nu_\tau}(E_{yy'},X'')}\nonumber\\ 
& & \times \exp \left[-\int_{X''}^{X'} dX'''/\lambda_\tau^{\rm dec}
       (E_y,X''')\right]
\end{eqnarray}
with $E_{yy'}=E_y/(1-y')=E/(1-y)(1-y')$.  Notice that the tau--flux 
$F_\tau$ is generated by the CC interactions of the initial $F_{\nu_\tau}^0$ flux
and attenuated in addition due to its decay.  In order to solve for 
$Z(E,X)$ iteratively as for the $\stackrel{(-)}{\nu}_\mu$ fluxes in the 
previous Section, one has to make a proper choice for the initial input. 
Due to the $D_{\nu_\tau}$ function in the exponential in (18)
with $D_{\nu_\tau}\to 0$ in the relevant $y\to 0$ region, the 
iterative result for $Z_{\nu_\tau}(E,X)$ is very robust with respect
to the initial input choice, as discussed after (7).  Therefore we use
again $Z_{\nu_\tau}^{(0)}(E,X')=0$ on the rhs of (18).  In the case of
$Z_\tau(E,X)$ in (19) there is, however, no equivalent exponential as
in (18) and thus the convergence of the iterative procedure becomes
sensitive to the input choice.  It turns out that a convenient and 
efficient input choice can be obtained by implementing the peculiar $E$
and $X$ dependence as implied by the $\tau$--decay contributions in (19)
from the very beginning.  This can be achieved by choosing a vanishing
$Z$--factor on the rhs of $Z_\tau$ in (19), in which case the
$X'$--integral can be performed analytically \cite{ref36} and the input
for the total $Z$--factor becomes \cite{ref13}
%\Eq.(21)
\begin{eqnarray}
Z^{(0)}(E,X) & = & \frac{\lambda_\nu(E)}{\lambda_\tau^{\rm dec}(E,\theta)}
 \int_0^1 dy \int_0^1 dy'\, K_\tau^{\rm dec}(E,y)\, K_\nu^{\rm CC}(E_y,y')\, 
  \lambda_\nu^{-1}\, (E_y)\, \eta_\nu(E,y)\, \eta_\nu(E_y,y')\nonumber\\
& & 
\times \frac{1}{XD_{\nu\tau}(E_y,E_{yy'})}\, 
\Big\{ \frac{1}{D_{\tau\nu}(E,E_y)}\left(1-e^{-XD_{\tau\nu}(E,E_y)}\right)
\nonumber\\
& &  
-\frac{1}{D_\nu(E,E_{yy'})} \left(1-e^{-XD_\nu(E,E_{yy'})}\right)\Big\}
\nonumber\\
& \simeq & \lambda_\nu(E)\int_0^1 \frac{dy}{1-y} \int_0^1 dy'\,
  K_\tau^{\rm dec}(E,y)\, K_\nu^{\rm CC}(E_y,y')\, \lambda_\nu^{-1}(E_y)\,
   \eta_\nu(E,y)\, \eta_\nu(E_y,y')\nonumber\\
& &
\times \frac{1}{XD_\nu(E,E_{yy'})}\, \left(1-e^{-XD_\nu(E,E_{yy'})}\right)
\end{eqnarray}
where the last approximation is due to 
$\lambda_\tau^{\rm dec}\ll \lambda_\nu$ in the relevant energy region
$E<10^8$ GeV, i.e., $Z^{(0)}$ becomes practically independent of the decay
length $\lambda_\tau^{\rm dec}$.  Furthermore $D_\nu(E,E_y)$ is given
in (9), 
$D_{\nu\tau}(E,E_y)=1/\lambda_\nu(E_y)-1/\lambda_\tau^{\rm dec}(E,\theta)$
and $D_{\tau\nu}(E,E_y)= -D_{\nu\tau}(E_y,E)$.  We have checked that this
input guarantees, for all cosmic neutrino fluxes considered at present,
a faster convergence of the iterations than choosing \cite{ref4} the 
solution for the $\nu_\mu$ flux as an input, $Z^{(0)}=Z_{\nu_\mu}^{(1)}$
with $Z_{\nu_\mu}^{(1)}$ given in (8).  Moreover, choosing \cite{ref10}
a vanishing initial input, $Z^{(0)}=0$, as was perfectly sufficient for
the $\nu_\mu$ fluxes, results in the worst, i.e., slowest convergence
of the iterative procedure.  One can now rewrite the solution for 
$Z(E,X)$ in (18) and (19) after the n--th iteration as \cite{ref36}
%Eq.(22)
\begin{eqnarray}
Z^{(n+1)}(E,X) & = & \frac{1}{X}\int_0^X dX'\int_0^1 dy\, K_\nu^{\rm NC}
  (E,y)\, \eta_\nu(E,y)\, e^{-X'D_{\nu_\tau}^{(n)}(E,E_y,X')}
\nonumber\\
& & 
+\, \frac{\lambda_\nu(E)}{\lambda_\tau^{\rm dec}(E,\theta)}\, \frac{1}{X}
 \int_0^X dX' \int_0^1 dy \, K_\tau^{\rm dec}(E,y)\, \eta_\nu(E,y)
      \lambda_\nu^{-1}(E_y)
\nonumber\\
& &
\times \, e^{-X'/\lambda_\tau^{\rm dec}(E_y,\theta)} \, 
         e^{X'/\Lambda_{\nu_\tau}^{(n)}(E,X')}
  \int_0^{X'} dX'' \int_0^1 dy'\, K_\nu^{\rm CC}(E_y,y')\, \eta_\nu(E_y,y')
\nonumber\\
& &
\times \, e^{-X''/\Lambda_{\nu_\tau}^{(n)}(E_{yy'},X'')}\, 
        e^{X''/\lambda_\tau^{\rm dec}(E_y,\theta)}
\end{eqnarray}
where $\Lambda_{\nu_\tau}^{(n)}(E,X') = \lambda_\nu(E)/[1-Z^{(n)}(E,X')]$,
i.e.,
%Eq.(23)
\begin{equation}
D_{\nu_\tau}^{(n)}(E,E_y,X') = \frac{1-Z^{(n)}(E_y,X')}{\lambda_\nu(E_y)}
 \, - \, \frac{1-Z^{(n)}(E,X')}{\lambda_\nu(E)}\,\, .
\end{equation}
Accordingly, the iterations have to be started with our initial $n=0$
input in (21).  After having obtained the final convergent result for
$Z^{(n+1)}$, the final $\nu_\tau$ flux $F_{\nu_\tau}^{(n+1)}(E,X)$ follows
from (16),
%Eq.(24)
\begin{equation}
F_{\nu_\tau}^{(n+1)}(E,X) = F_{\nu_\tau}^0(E)\, 
    e^{-X/\Lambda_{\nu_\tau}^{(n+1)}(E,X)}  \,\,  ,
\end{equation}
which in turn gives the $\tau$--flux
%Eq.25
\begin{equation}
F_\tau^{(n+1)}(E,X) = \frac{1}{\lambda_\nu(E)}\, 
  e^{-X/\lambda_\tau^{\rm dec}(E,\theta)} 
   \int_0^X dX' \int_0^1 \frac{dy}{1-y}\, K_\nu^{\rm CC}(E,y)\, 
    F_{\nu_\tau}^{(n+1)}(E_y,X')\,
     e^{X'/\lambda_\tau^{\rm dec}(E,\theta)}\, \, .
\end{equation}
Similar expressions hold for antineutrinos as well.

The iterative results for the total $Z$--factor in (17) are shown in
Figs.~4 and 5 where the initial input $Z^{(0)}$, as given in (21), is
displayed by the dotted curves.  For the generic initial $E^{-1}$ and
$E^{-2}$ fluxes in (10) and (11) we also show in Fig.~4 the results 
after the third iteration, $Z^{(3)}$, in order to illustrate the rate
of convergence as well as its dependence on the nadir angle 
$\theta=0^{\rm o}$ ($X=1.1\times 10^5$ km we) and $\theta = 50^{\rm o}$
($X = 3.6\times 10^4$ km we).  In general it turns out that already the
second ($n=2$) iteration yields sufficiently accurate results, $Z^{(2)}$,
provided one uses as input $Z^{(0)}$ in (21) as implied by the 
$\tau$--decay.  This holds also for the rather hard initial $E^{-1}$
flux in Fig.~4 and the AGN--SS flux in Fig.~5 which imply large
$Z$--factors, $Z\gg 1$.  This is so because the maximum difference
between the results of the next $n=3$ iteration $Z^{(3)}$ and $Z^{(2)}$
is less than about 5\% for all relevant initial cosmic neutrino fluxes.
An accuracy of less than about 5\% is certainly sufficient in view of
the uncertainties inherent to models of cosmic neutrino fluxes
(cf.~Fig.~1).  Obviously the iterative convergence improves even more
for larger values of $\theta$, i.e., smaller depths $X$, as can be
deduced from Fig.~4.  It should be emphasized that, in contrast to the 
case of muon neutrinos in Sec.~2, the {\em first} $n=1$ iterative
results for $Z^{(1)}$ are {\em not} sufficiently accurate as can be 
seen from Figs.~4 and 5 by comparing the dashed curves ($Z^{(1)}$) with
the solid ones ($Z^{(2)}$): in some cases (harder initial fluxes) 
$Z^{(2)}$ becomes larger than $Z^{(1)}$ by about 20\%.  The results 
for $\bar{\nu}_\tau$, $Z=Z_{\bar{\nu}_\tau}+Z_{\tau^+}$, are again
similar but larger than for $\nu_\tau$, $ Z=Z_{\nu_\tau}+Z_{\tau^-}$, 
for $E$
\raisebox{-0.1cm}{$\stackrel{<}{\sim}$} 10$^6$ GeV where 
$\lambda_{\bar{\nu}}>\lambda_\nu$.  Inserting the various iterative
solutions of Figs.~4 and 5 into (16) we obtain the $\nu_\tau$ fluxes
for a given n--th iteration, $F_{\nu_\tau}^{(n)}(E,X)$.  The ratios
of these fluxes for two consecutive iterations, 
$F_{\nu_\tau}^{(n+1)}/F_{\nu_\tau}^{(n)}$, are displayed in Figs.~6 and
7.  Whereas the first iteration relative to the zeroth input,
$F_{\nu_\tau}^{(1)}/F_{\nu_\tau}^{(0)}$, is way off the final result
as shown by the dashed curves, the second iteration suffices already
for obtaining a sufficiently accurate result as illustrated by
$F_{\nu_\tau}^{(2)}/F_{\nu_\tau}^{(1)}$ by the solid curves. This is
supported by the fact that an additional third iteration results in
$|F_{\nu_\tau}^{(3)}/F_{\nu_\tau}^{(2)}-1|<0.05$ for all relevant
initial cosmic neutrino fluxes considered.  
(This stability does {\em not} hold \cite{ref7} for initial fluxes
$F_{\nu_\mu +\bar{\nu}_\mu}^0\sim E^{-1}$ without an appropriate 
$E^{-2}$ cutoff in (10) at very high energies,
or for fluxes which are partly even flatter than $E^{-1}$
up to highest energies like the $Z$--burst flux in Fig.~1. This
instability is caused by the fact that $\eta_\nu(E,y) =1$ for
$F_\nu^0\sim E^{-1}$ in $Z_\tau$ in (19); thus the huge spike of
$d\sigma_{\nu N}^{\rm CC}/dy$ at $y\to 1$ in (19) and (20) does not
get damped by powers of $(1-y)$ -- as opposed to, e.g., $F_\nu^0
\sim E^{-2}$ where $\eta_\nu= 1-y$.  This, however, is of no
concern for $Z_{\nu_\mu}$ and $Z_{\nu_\tau}$ in (5) and (18),
respectively, since there the integrands are exponentially 
suppressed as $y\to 1$ via $\exp\left[ -X'D_\nu(E,E_y,X)\right]$.
Notice that the $Z$--burst flux is far too small for being tested
with upward--going muon events \cite{ref5,ref13}.)
Therefore we consider 
$F_{\stackrel{(-)}{\nu}_\tau}^{(2)}(E,X)$ as our final result.
It is furthermore obvious from Figs.~6 and 7 that the convergence of
the iterative procedure strongly improves for increasing values of
$\theta$ (decreasing $X$) as illustrated for $\theta =50^{\rm o}$.

The resulting total $\nu_\tau+\bar{\nu}_\tau$ fluxes for the various
initial total cosmic fluxes $F_{\nu_\tau+\bar{\nu}_\tau}^0(E)$ are
shown in Figs.~8 and 9 for three typical nadir angles $\theta=0^{\rm o}$
($X=1.1\times 10^5$ km we), $\theta =30^{\rm o}$ ($X=6.8\times 10^4$
km we) and $\theta =60^{\rm o}$ ($X=2.6 \times 10^4$ km we).  The 
typical enhancement (`bump') of the attenuated and regenerated
$\stackrel{(-)}{\nu}_\tau$ flux around $10^4$ -- $10^5$ GeV at small
values of $\theta$, which is prominent for harder (flatter) initial
fluxes like $F_{\stackrel{(-)}{\nu}_\tau}^0\sim E^{-1}$ in Fig.~8, and
which is absent for $\stackrel{(-)}{\nu}_\mu$ fluxes, agrees with the
original results of \cite{ref4,ref7,ref9}, as was also confirmed
by a Monte Carlo simulation \cite{ref34}.  Such an enhancement is less
pronounced for the GRB--WB and TD--SLBY fluxes in Fig.~9, and is absent
for steeper fluxes like for the $E^{-2}$ one in Fig.~8 and for the even
steeper AGN--M95 flux in Fig.~9.  Regeneration is responsible for
an even more pronounced enhancement below $10^4$ GeV for the AGN--SS
flux in Fig.~9 since this flux is particularly hard below $10^5$ GeV
(cf.~Fig.~1). This latter result is shown mainly for theoretical
curiosity.  From now on we shall disregard the cosmic AGN--SS flux
since it is in serious conflict with recent experimental upper bounds
\cite{ref20,ref21} as can be seen in Fig.~1.  The results for the 
absolute total $\nu_\tau+\bar{\nu}_\tau$ and $\tau^-+\tau^+$ fluxes,
arising from the initial cosmic $\nu_\tau+\bar{\nu}_\tau$ fluxes,
are presented in Figs.~10 and 11.  The $\nu_\tau+\bar{\nu}_\tau$ 
results correspond of course to the relative ratios shown in Figs.~8
and 9.  Besides the generic initial fluxes in (10) and (11), we have
in addition used only those initial cosmic fluxes in Fig.~1 which give
rise to large enough upward--going $\mu^- +\mu^+$ event rates 
\cite{ref5,ref13} measurable in present and future experiments.  Note
that the $\tau^- +\tau^+$ fluxes in Figs.~10 and 11 at the detector
site, despite being (superficially) suppressed with respect to the
$\nu_\tau +\bar{\nu}_\tau$ fluxes, sizeably contribute to the 
upward--going $\mu^- +\mu^+$ and shower event rates \cite{ref13}.  
This is due to the fact that the $\tau$ fluxes do not require additional
weak interactions for producing $\mu$--events in contrast to the 
$\nu_\tau$ fluxes. Because of the prompt $\tau^{\pm}$ decays, they
furthermore give rise to a sizeable secondary $\bar{\nu}_\mu+\nu_\mu$
flux contribution to the original cosmic $\nu_\mu+\bar{\nu}_\mu$ flux,
which will be discussed in Sec.~4.

In \cite{ref13} a semi--analytic solution of the coupled transport
equations (12) and (13) has been presented and used, which was obtained
from the first $n=1$ iteration starting with a vanishing input 
$Z^0(E,X)=0$, instead of using (21).  As we have seen, this approach
does not provide sufficiently accurate results, despite opposite
claims in the literature \cite{ref10} (the first $n=1$ iteration is
sufficient only for very large values of $\theta$ close to $90^{\rm o}$,
i.e., very small values of $X/\rho={\cal{O}}$(100 km), relevant for
neutrinos skimming the Earth's crust).  This $n=1$ iterative solution
of \cite{ref13} underestimates the correct results in some extreme
cases (like for the hard initial $E^{-1}$ flux at $\theta=0^{\rm o}$)
by as much as 40\%.  On the other hand, for increasing values of 
$\theta$ this discrepancy disappears very quickly.  Consequently,
some of the {\em total} nadir--angle--integrated upward--going
$\mu^- +\mu^+$ event rates calculated in \cite{ref13} will be increased
by less than about 2\%.  This is due to the fact that 80\% of the 
$\mu^- +\mu^+$ rates are initiated by the $\nu_\mu+\bar{\nu}_\mu$
flux and only about 20\% derives from the $\nu_\tau +\bar{\nu}_\tau$
and the associated $\tau^- +\tau^+$ fluxes.  For completeness we 
present in Table 1 the correct expectations for the total $\mu^-+\mu^+$
event rates for the relevant dominant initial cosmic fluxes in Fig.~1,
using Eqs.~(12) and (14) of \cite{ref13} for calculating the rates
initiated by the $\nu_\tau +\bar{\nu}_\tau$ and $\tau^- +\tau^+$ fluxes, 
respectively.

Finally, it is also of interest to compare the tau--lepton range as
given by our semi--analytic approach of treating the energy loss 
continuously in (13), with the one obtained by a stochastic treatment
of the lepton energy loss (where the $\gamma(E)$ term in (13) is 
absent, i.e., the energy loss is treated separately, and the relevant
survival probability $P(E,X)$ is calculated using Monte Carlo
simulations, e.g., \cite{ref32,ref35,ref37,ref38}).  To do this, we
can drop the inhomogeneous neutrino term in (13) and the resulting
homogeneous transport equation for $F_\tau(E,X)$ can be easily solved
\cite{ref13}:
%Eq.(26)
\begin{equation}
F_\tau(E,X) = F_\tau(\bar{E}(X,E),0)\exp\left[ -\int_0^X 
    A(\bar{E}(X',E))\, dX'\right]
\end{equation}
with $A(E)\equiv 1/\hat{\lambda}(E)-\partial\gamma(E)/\partial E$,
$\hat{\lambda}^{-1}\equiv(\lambda_\tau^{\rm CC})^{-1}+
(\lambda_\tau^{\rm dec})^{-1}$, and where $d\bar{E}(X,E)/dX=
\gamma(\bar{E})$ with $\bar{E}(0,E)=E$.  The survival probability
$P(E_0,X)$ for a tau--lepton with an initial energy $E_0$ at $X=0$
is then defined by the ratio of the energy--integrated differential
fluxes $F_\tau$ at $X$ and $X=0$ :  assuming, as usual, a monoenergetic
initial flux in (26), $F_\tau(\bar{E}(X,E),0)\sim\delta(E-E_0)$, one
obtains \cite{ref10}
%Eq.(27)
\begin{equation}
P(E_0,X) = \frac{\gamma(\tilde{E}_0)}{\gamma(E_0)}\, \exp
 \left[-\int_0^X A(\tilde{E}_0(X',E_0))\, dX'\right]
\end{equation}
where we have used \cite{ref13} $d\bar{E}/dE =\gamma(\bar{E})/\gamma(E)$ 
and $d\tilde{E}_0(X,E_0)/dX=-\gamma(\tilde{E}_0)$ with 
$\tilde{E}_0(0,E_0)=E_0$.  The (tau) lepton range for an incident
lepton energy $E$ and a final energy $\tilde{E}(X,E)$ required to be
greater than $E^{\rm min}$ at the detector, say, is then defined by
%Eq,(28)
\begin{equation}
R(E) = \int_0^{X_{\rm max}} P(E,X)\,  dX
\end{equation}
where we have substituted $E$ for $E_0$ in (27) and the upper limit
of integration $X_{\rm max}$ derives from 
$\tilde{E}(X,E)\geq E^{\rm min}$. (Notice that for energy--independent
values of $\alpha$ and $\beta$ in $\gamma(E)=\alpha+\beta E$ one
simply gets $X_{\rm max} = \frac{1}{\beta}\ln \frac{\alpha+\beta E}
{\alpha +\beta E^{\rm min}}$.) For calculating the $\tau$--lepton
range $R_\tau(E)$ we use in $\gamma_\tau$ for the ionization energy
loss \cite{ref39,ref40} $\alpha_\tau \simeq 2.0 \times 10^{-3}$ GeV
(cm we)$^{-1}$ and for the radiative energy loss (through 
bremsstrahlung, pair production and photonuclear interactions)
\cite{ref10} $\beta_\tau=\beta_\tau(E)\simeq[0.16+0.6(E/10^9$ 
GeV)$^{0.2}] \times 10^{-6}$ (cm we)$^{-1}$ which parametrizes 
explicit model calculations \cite{ref32,ref35} for standard rock 
($\rho=2.65$ g/cm$^3$) reasonably well for 
$10^3$ \raisebox{-0.1cm}{$\stackrel{<}{\sim}$} $E$ 
\raisebox{-0.1cm}{$\stackrel{<}{\sim}$} $10^9$ GeV.
Furthermore we impose \cite{ref32} $E^{\rm min}=50$ GeV.  Our results
for the $\tau$--lepton range are shown in Fig.~12 which agree of course
with the ones in \cite{ref10}.  The $\tau$--decay term dictates the
$\tau$--range until $E>10^7$ GeV where the tau--lepton energy loss
becomes relevant.  The dashed--dotted curve shows the range as obtained
by omitting the contribution due to the CC interaction length 
$\lambda_\tau^{\rm CC}$ in $\hat{\lambda}$ in (27).  This term will
be relevant for 
$E$ \raisebox{-0.1cm}{$\stackrel{>}{\sim}$} $10^{10}$ GeV where
$\lambda_\tau^{\rm CC}$ becomes comparable to $\lambda_\tau^{\rm dec}$
as evident from Fig.~13.  For comparison, stochastic Monte Carlo
evaluations \cite{ref32,ref35,ref38} of the $\tau$--range are shown in
Fig.~12 by the dotted curves which are of course strongly dependent
on the assumed model extrapolations of the radiative cross sections
to ultrahigh energies.  Our results depend obviously also on such
extrapolations due to specific choices of $\beta_\tau(E)$.
Nevertheless, one concludes \cite{ref10} from Fig.~12 that the 
{\em continuous} tau--lepton energy loss approach, as used in (13),
yields very {\em similar} results as the stochastic Monte Carlo
calculations where the energy loss is treated separately. 

A similar conclusion holds for the muon--range $R_\mu(E)$ which we
show for completeness in Fig.~14.  Within the continuous muon energy
loss approach, $R_\mu$ follows from (28) and (27) where the
$\lambda_\tau^{\rm dec}$ term has to be omitted and in 
$\gamma_\mu(E)=\alpha_\mu+\beta_\mu E$ we take $\alpha_\mu\simeq
\alpha_\tau\simeq 2.0\times 10^{-3}$ GeV (cm we)$^{-1}$ and 
\cite{ref32,ref35} $\beta_\mu\simeq 6.0\times 10^{-6}$ (cm we)$^{-1}$
which, moreover, reproduces best \cite{ref5} the
Monte Carlo result of Lipari and Stanev \cite{ref37}
for the average muon--range in standard rock for $E>10^3$ GeV.
Furthermore we choose \cite{ref32} the final muon energy to be 
larger than $E^{\rm min}=1$ GeV.  
(It should be noted that here $P_\mu(E,X)\simeq 1$ in (27), i.e.,
$R_\mu(E)\simeq X_{\rm max}$ in (28).)  The muon--range $R_\mu$
calculated within the continuous muon energy loss approach yields
again, as in the case of taus, very similar results as the stochastic
Monte Carlo calculations \cite{ref32,ref35,ref38} for 
$R_\mu^{\rm sto}$ as shown in Fig.~14.  This is contrary to the
conclusions reached in \cite{ref10} that the continuous approach to
the muon energy loss overestimates the muon--range as compared to
stochastic Monte Carlo simulations.  Therefore the continuous
approach to the lepton energy loss is applicable to {\em both}
taus and muons, since in both cases it yields similar results for
the lepton ranges as the stochastic Monte Carlo simulations with
the energy loss being treated separately.
 
\section{The transport equation of muon neutrinos including 
secondary muon neutrinos from tau neutrino interactions}

It has been pointed out \cite{ref14} that the $\nu_\tau -\tau$
regeneration chain $\nu_\tau\to\tau\to\nu_\tau\to\ldots$ creates a
secondary $\bar{\nu}_\mu +\nu_\mu$ flux due to the prompt, purely
leptonic, tau decays $\tau^-\to\nu_\tau\mu^-\bar{\nu}_\mu$ and
$\tau^+\to\bar{\nu}_\tau\mu^+\nu_\mu$.  This will enhance the 
regenerated $\stackrel{(-)}{\nu}_\mu$ fluxes calculated according 
to (1) and
thus also the `naively´' calculated \cite{ref5,ref15,ref41} 
upward--going muon event rates at the detector site.  Secondary
neutrinos originate from 
the associated 
$\tau^{\pm}$ flux $F_\tau(E,X)$ and a prompt $\tau$--decay like
$\tau^-\to\bar{\nu}_\mu X'$. Adding those contributions,  denoted by 
$G_{\tau^{\stackrel{(-)}{+}}\to\stackrel{(-)}{\nu}_\mu}(E,X)$,
to the simple transport equation (1) used
thus far one obtains
%Eq.(29)
\begin{equation}
\frac{\partial F_{\nu_\mu}(E,X)}{\partial X} =
-\frac{F_{\nu_\mu}(E,X)}{\lambda_\nu(E)}\, +\, 
  \frac{1}{\lambda_\nu(E)}
   \int_0^1\frac{dy}{1-y}\, K_\nu^{\rm NC}(E,y)\,
    F_{\nu_\mu}(E_y,X)+G(E,X)
\end{equation}
with $G=G_{\tau^+\to\nu_\mu}$ where
%Eq.(30)
\begin{equation}
 G_{\tau^+\to\nu_{\mu}}(E,X) =
  \frac{1}{\lambda_\tau^{\rm dec}(E,\theta)}
   \int_0^1\frac{dy}{1-y}\, K_{\tau^+}^{\rm dec}(E,y)\, 
    F_{\tau^+}(E_y,X) 
\end{equation}
and a similar transport equation holds
for $F_{\bar{\nu}_\mu}$ with an appropriate expression for 
$G_{\tau^-\to\bar{\nu}_\mu}$.
The relevant $\tau$ fluxes
$F_{\tau^{\pm}}$ have been calculated in the previous
Section (cf.~Figs.~10 and 11).
As in (14), the decay kernel in (30) is 
$K_{\tau^+}^{\rm dec}(E,y) = (1-y)\, dn_{\tau^+\to\nu_\mu}(z)/dy$
with $z=1-y$ and
the relevant $\tau^+\to\nu_\mu X'$ decay distribution is 
given by \cite{ref28}
%Eq.(31)
\begin{equation}
\frac{dn_{\tau^+\to\nu_\mu}(z)}{dz} = B_{\nu_\mu}
 \left[2-6z^2+4z^3+P(-2+12z-18z^2+8z^3)\right]
\end{equation}
with $P=+1$ and the branching fraction $B_{\nu_\mu}=0.18$. For a
decaying $\tau^-\to\bar{\nu}_\mu X'$ one has $P=-1$ in (31). Notice
that the $\stackrel{(-)}{\nu}_\mu$ spectrum in (31) is a little
softer than the $\stackrel{(-)}{\nu}_\tau$ spectrum from the 
$\tau^{\pm}\to\stackrel{(-)}{\nu}_\tau X'$ decay \cite{ref7,ref28} in
(15).   

It should be noticed that the contribution of secondary neutrinos may
alternatively be calculated using directly the 
$\stackrel{(-)}{\nu}_\tau$ fluxes $F_{\stackrel{(-)}{\nu}_\tau}(E,X)$
which give rise to the reaction chains 
$\nu_\tau\stackrel{\rm CC}{\longrightarrow}\tau^- \to 
\bar{\nu}_{\mu}X'$ and
$\bar{\nu}_\tau\stackrel{\rm CC}{\longrightarrow}\tau^+
\to \nu_{\mu}X'$.
Denoting these contributions by 
$G_{\nu_{\tau}\to\bar{\nu}_\mu}(E,X)$ and
$G_{\bar{\nu}_\tau \to \nu_\mu}(E,X)$, respectively,
the inhomogeneous term in the transport equation (29) is given by
$G=G_{\bar{\nu}_\tau \to\nu_\mu}$ with
%Eq.(32)
\begin{equation}
G_{\bar{\nu}_\tau\to\nu_{\mu}}(E,X) = N_A
  \int^1_0 \frac{dy}{1-y}\int_0^1\frac{dz}{z}\, 
   \frac{dn_{\tau^+\to\nu_\mu}(z)}{dz}\, 
    \frac{d\sigma_{\bar{\nu}N}^{\rm CC}(\frac{E_y}{z},y)}{dy}\, 
     F_{\bar{\nu}_\tau}\left(\frac{E_y}{z},\, X\right)
\end{equation}
where $E_y/z=E/(1-y)z$, the decay distribution is given by (31) and the 
relevant flux $F_{\bar{\nu}_\tau}$ has been calculated in the previous
Section (cf.~Figs.~10 and 11).  Although (32) and (30) yield the same
quantitative results for $F_{\nu_\mu}(E,X)$, these two expressions should
{\em not} be added since it would correspond to double--counting the 
effect of secondary neutrino production.  This is due to the fact that
the CC contribution $G_{\nu_{\tau}\to \tau}$ has been already included
in (13) [third term on the rhs] for calculating $F_\tau$.  (The situation
here is very similar to the calculation of the atmospheric muon flux
\cite{ref31,ref28} where almost all muons come from meson decays with
the meson flux being generated by nucleon interactions with air, i.e.,
by nucleon $\to$ meson transitions.  These latter transitions are taken
into account only in the evolution equation of the meson flux, but not
anymore for the muon flux evolution.)  For definiteness, we use the 
simpler expression in (30) for our subsequent calculations.

As in our previous cases, the most general transport equation (29)
for muon neutrinos is easily solved by an ansatz like (16) for tau
neutrinos,
%Eq.(33)
\begin{equation}
F_{\nu_\mu}(E,X) = F_{\nu_\mu}^0(E)\exp\, 
 \left[ -\frac{X}{\Lambda_{\nu_{\mu}G}(E,X)}\right]
\end{equation}
with
%Eq.(34)
\begin{equation}
\Lambda_{\nu_\mu G}(E,X) = 
  \frac{\lambda_\nu(E)}{1-Z_{\nu_{\mu}G}(E,X)}
\end{equation}
and $Z_{\nu_{\mu}G}=Z_{\nu_\mu}+Z_G$.  Inserting (33) into (29) one
obtains 
%Eq.(35)
\begin{equation}
Z_{\nu_\mu}(E,X) =\frac{1}{X}\int_0^X dX' \int dy\, 
 K_\nu^{\rm NC}(E,y)\, \eta_\nu(E,y)\, e^{-X'D_{\nu_\mu}(E,E_y,X')}
\end{equation}
which is similar to (5) but with 
$D_{\nu_\mu}(E,E_y,X')=\Lambda_{\nu_\mu G}^{-1}(E_y,X')
  -\Lambda_{\nu_\mu G}^{-1}(E,X')$, and
%Eq.(36)
\begin{equation}
Z_G(E,X) = \frac{\lambda_\nu(E)}{F_\nu^0(E)}\, \frac{1}{X}\,
 \int_0^X dX'\, G(E,X')\, e^{X'/\Lambda_{\nu_\mu G}(E,X')}\, \, .
\end{equation}
Using again an iteration algorithm to solve for 
$Z_{\nu_\mu G}(E,X)$, the solution of (35) and (36) after the n--th
iteration becomes
%Eq.(37)
\begin{eqnarray}
Z_{\nu_\mu G}^{(n+1)}(E,X) & = & \frac{1}{X}\int_0^X dX'\int_0^1 dy\, 
  K_\nu^{\rm NC}(E,y)\, \eta_\nu(E,y)\, 
   e^{-X'D_{\nu_\mu}^{(n)}(E,E_y,X')}
\nonumber\\
& & +\frac{\lambda_\nu(E)}{F_{\nu_\mu}^0(E)} \, \frac{1}{X}\, 
  \int_0^X dX' G(E,X')\, e^{X'/\Lambda_{\nu_\mu G}^{(n)}}(E,X')
\end{eqnarray}
where $D_{\nu_\mu}^{(n)}$ is   as   defined   above with 
$\Lambda_{\nu_\mu G}\!\!\!\!\to\!\!\!\!\Lambda_{\nu_\mu G}^{(n)}$ 
and
$\Lambda_{\nu_\mu G}^{(n)}(E,X')=\lambda_\nu(E)/$
$\left[1-Z_{\nu_\mu G}^{(n)}(E,X')\right]$.
Due to the dominant and large $\tau$--decay contribution $G(E,X)$ 
in (29), it turns out that the optimal input choice for providing
sufficiently convergent iterative solutions is obtained by 
implementing, as in the case of tau neutrinos in Sec.~3, the 
peculiar $E$ and $X$ dependence as implied by the $\tau$--decays, 
i.e., by $G$ in (36) from the very beginning.  Therefore we use
again (see discussion after Eq.~(20)) $Z_{\nu_\mu}^{(0)}(E,X)=0$
and a vanishing $Z$--factor on the rhs of $Z_G$ in (36) which 
gives for the total input $Z$--factor
%Eq.(38)
\begin{equation}
Z_{\nu_\mu G}^{(0)}(E,X) = \frac{\lambda_\nu(E)}{F_\nu^0(E)}\, 
  \frac{1}{X} \int_0^X dX'\, G(E,X')\, e^{X'/\lambda_\nu(E)}\,\,.
\end{equation}
Inserting this into the rhs of (37) results in the first iterative
solution $Z_{\nu_\mu G}^{(1)}(E,X)$, and so on.  In contrast to 
$Z_{\nu_\mu}^{(1)}$ in (8), $Z_{\nu_\mu G}^{(1)}$ does not 
provide us with a sufficiently accurate final result, i.e., the 
maximum difference between $Z_{\nu_\mu G}^{(1)}$ and 
$Z_{\nu_\mu G}^{(2)}$ is here {\em not} always less than about 5\%
for some initial cosmic neutrino fluxes and energies.  Therefore
we have to carry out one further iteration, as in the case of
tau neutrinos in Sec.~3, by inserting $Z_{\nu_\mu G}^{(1)}$ into
the rhs of (37) in order to obtain $Z_{\nu_\mu G}^{(2)}(E,X)$
which turns out to be sufficiently close to the final result since
$|Z_{\nu_\mu G}^{(3)}/Z_{\nu_\mu G}^{(2)}-1|$
\raisebox{-0.1cm}{$\stackrel{<}{\sim}$} 0.02.

Our iterative results for $Z_{\nu_\mu G}^{(1,2)}$ are shown in
Figs.~15 and 16 together with the appropriate input 
$Z_{\nu_\mu G}^{(0)}$ in (38) shown by the dotted curves.
In order to illustrate the faster iterative convergence for increasing
$\theta$ (smaller $X$), the results for $\theta=50^{\rm o}$ are 
presented in Fig.~15 as well.  The sufficiently accurate results
$Z_{\nu_\mu G}^{(2)}(E,X)$ and the similar expressions for 
$Z_{\bar{\nu}_\mu G}^{(2)}$, when inserted into (33), yield the
final total fluxes $F_{\nu_\mu +\bar{\nu}_\mu}(E,X)$ shown in
Figs.~17 and 18.  The effect and importance of secondary neutrinos
is best seen by comparing our results (solid and dashed curves)
with the usual ones \cite{ref3,ref5,ref7,ref15} obtained just for
primary muon neutrinos ($G\equiv 0$ in (29)) shown by the dotted
curves, which correspond of course to the results obtained in
Sec.~2.  Our results in Fig.~17 agree with the ones obtained in
\cite{ref9}, within the approximations made there.

The corresponding $\stackrel{(-)}\nu_\mu$ initiated upward--going
$\mu^{\stackrel{(+)}{-}}$ event rate per unit solid angle and second
is calculated according to 
%Eq.(39)
\begin{equation}
N_{\mu^-}^{(\nu_\mu)} = N_A \int_{E_{\mu}^{\rm min}} dE_\nu
 \int_0^{1-E_\mu^{\rm min}/E_\nu} dy \, A(E_\mu)\, 
   R_\mu(E_\mu,E_\mu^{\rm min})
    \frac{d\sigma_{\nu_\mu N}^{\rm CC}\,(E_\nu,y)}{dy}\, 
      F_{\nu_\mu}(E_\nu,X)
\end{equation}
with $E_\mu=(1-y)E_\nu$ and the energy dependent area $A(E_\mu)$
of the considered underground detectors is taken as summarized in
\cite{ref5}.  The muon--range is given by 
$R_\mu(E_\mu,E_\mu^{\rm min}) =\frac{1}{\beta_\mu}\ln \frac
{\alpha_\mu+\beta_\mu E_\mu}{\alpha_\mu+\beta_\mu E_\mu^{\rm min}}$.
It describes the range of an energetic muon being produced with
energy $E_\mu$ and, as it passes through the Earth loses energy,
arrives at the detector with energy above $E_\mu^{\rm min}$. 
The energy--loss parameters are taken as at the end of the previous 
Section, i.e., $\alpha_\mu = 2\times 10^{-3}$ GeV (cm we)$^{-1}$ and
$\beta_\mu = 6\times 10^{-6}$ GeV (cm we)$^{-1}$.  The integral
over the neutrino energy $E_\nu$ was, for definiteness and better
comparison \cite{ref9}, performed up to a maximum neutrino energy
of 10$^8$ GeV.  The differential $\theta$--dependent $\mu^- +\mu^+$
rates for $E_\mu^{\rm min}=10^4$ GeV and $E_\mu^{\rm min}=10^5$
GeV are shown in Figs.~19 and 20. We also include the contributions
initiated by the primary $\nu_\mu +\bar{\nu}_\mu$ flux, for brevity
denoted by $\nu_\mu\to\mu$, and by the $\nu_\tau+\bar{\nu}_\tau$
flux via $\nu_\tau\to\tau\to\mu$ and the $\tau^- +\tau^+$ flux 
via $\tau \to\mu$ as discussed in the previous Section.  The
secondary neutrino contributions to the muon event rates have
their largest relative contributions obviously at small nadir
angles, with an enhancement over the primary 
$\nu_\mu+\nu_\tau+\tau$ initiated rates of up to 40\% for the 
hard $E^{-1}$, AGN--M95 and TD--SLBY initial fluxes. 

At small nadir angles, however, the event rates are smallest and
statistics are low.  For \mbox{$\theta$ 
\raisebox{-0.1cm}{$\stackrel{>}{\sim}$} 60$^{\rm o}$} the event
rates are roughly a factor of (more than) 10 larger and the 
enhancement of the overall $\nu_\mu+\nu_\tau+\tau$ initiated muon
rates (dashed--dotted curves in Figs.~19 and 20) can not be larger
than about 15\%.  These results are more explicitly illustrated in 
Tables 2 and 3 where we present, besides the total 
nadir--angle--integrated rates, also the ones integrated over 
three typical $\theta$--intervals.  (Remember that this amounts
to integrating (39) over 
$\int_0^{2\pi} d\varphi \int_{\theta_{\min}}^{\theta_{\rm max}} 
d\theta\sin\theta =
2\pi\int_{\theta_{\rm min}}^{\theta_{\rm max}} d\theta\sin\theta$, 
with
$\theta_{\rm min}=0^{\rm o}$ and
$\theta_{\rm max}=90^{\rm o}$ for the total rates.)  Since 
secondary muon neutrinos contribute significantly to a muon 
excess only at small and medium nadir angles, $\theta<60^{\rm o}$,
the primary event rates (shown in brackets in Tables 2 and 3) can
be enhanced by more than 20\%, in particular for $E_\mu^{\rm min}=
10^5$ GeV.  The statistics, however, are low since the fluxes are
already strongly attenuated for $\theta<60^{\rm o}$ (large $X$),
cf.~Figs.~17 and 18.  On the other hand, most of the events are
generated at large $\theta$, $\theta>60^{\rm o}$, where the
effect of secondary neutrinos is sizeably reduced (cf.~Figs.~19
and 20), the total rates in Tables 2 and 3 are increased by 
less than 10\%.  Since the expected angular resolution of present
and proposed detectors \cite{ref21,ref42} is typically about
$1^{\rm o}/(E_\nu/$TeV)$^{0.7}$, differential $\theta$--dependent 
measurements should be feasible, in order to delineate
experimentally
the effects of secondary neutrino fluxes.  Keeping in mind that 
the lifetime 
of the planned experiments is roughly ten years, it 
appears to be not unreasonable that the tenfold rates implied by
Tables 2 and 3 may be observable in the not too distant future.

\section{Summary and Conclusions}

For the sake of completeness we have first studied the solutions
of the single transport equation for cosmic 
$\stackrel{(-)}{\nu}_\mu$ neutrinos propagating through the Earth.
Although frequently used, the excellent convergence of its 
iterative solutions has not been explicitly demonstrated thus far
for more realistic and hard cosmic neutrino fluxes.  Using the 
symbolic ansatz for the solution $F_\nu(E,X)= F_\nu^{0}(E)\exp
\left[(1-Z)X/\lambda_\nu\right]$, with $\lambda_\nu$ being the 
neutrino interaction length, the most simple input choice
$Z_{\nu_\mu}^{(0)}(E,X)=0$ suffices to produce a sufficiently
accurate iterative result $Z_{\nu_\mu}^{(1)}$ already after the
first iteration, for all presently used initial cosmic model
fluxes $F_\nu^0$.

Turning to the iterative solutions of the far more complicated
coupled transport equations for $\stackrel{(-)}{\nu}_\tau$ and 
their associated $\tau^{\pm}$ fluxes, a new semi--analytic input
algorithm is presented which allows for a fast convergence of
the iterative solutions:  already a second $n=2$ iteration suffices
for obtaining a sufficiently accurate result $Z^{(2)}$ and thus
for the final $F_{\stackrel{(-)}{\nu}_\tau}$ and its associated
$F_{\tau^{\pm}}$ fluxes.  In order to achieve this one has to
implement the peculiar $E$ and $X$ dependence as implied by the
$\tau^{\pm}$ decay contributions already into the initial zeroth
order input $Z^{(0)}(E,X)$. Choosing a vanishing input $Z^{(0)}=0$
as in the case for $\stackrel{(-)}{\nu}_\mu$ fluxes or even the
final solution for the $\stackrel{(-)}{\nu}_\mu$ flux as an input,
$Z^{(0)}=Z_{\nu_\mu}^{(1)}$, as frequently done, results in a 
far slower convergence of the iterative procedure.  For
completeness we briefly outline also the implications for the
upward--going $\mu^- +\mu^+$ event rates for underground neutrino
detectors using some relevant cosmic neutrino fluxes.  These
events are generated by the so called  `primary'
$\stackrel{(-)}{\nu}_\mu$, $\stackrel{(-)}{\nu}_\tau$ and 
$\tau^{\pm}$ fluxes via the weak transitions and decays 
$\nu_\mu\stackrel{\rm CC}{\to}\mu$, 
$\nu_\tau\stackrel{\rm CC}{\to}\tau\to\mu$ and $\tau\to\mu$.
Furthermore, for calculating the range $R_\tau(E)$ of tau--leptons,
their energy loss can either be treated `continuously' by 
including it directly in the transport equation, or `stochastically'
by treating it separately.  Both approaches give very similar 
results for $R_\tau$ up to highest energies of $10^{12}$ GeV
relevant at present.  A similar agreement is obtained for the 
muon range $R_{\mu}(E)$.  Therefore the continuous approach is
applicable to both taus and muons.  This is contrary to claims
in the literature that the continuous approach overestimates
$R_\mu$ as compared to stochastic Monte Carlo simulations.

Finally, we generalized the single transport equation for 
$\stackrel{(-)}{\nu}_\mu$, by taking into account the contributions
of secondary $\nu_\mu$ and $\bar{\nu}_\mu$ fluxes.  These so called
`secondary' muon neutrino fluxes originate from prompt $\tau^{\pm}$
decays where the $\tau$--leptons are generated by the regeneration
chain $\nu_\tau\to\tau\to\nu_\tau\to\ldots$ when a cosmic 
$\nu_\tau$ passes through the Earth.  Thus the 
secondary $\nu_\mu +\bar{\nu}_\mu$ flux arises from the associated 
$\tau^{\pm}$ flux, as obtained from the coupled transport 
equations for $\nu_\tau$ and $\tau$, which initiates the $\tau\to
\nu_\mu$ transitions ($\tau^-\to\nu_\tau\mu^-\bar{\nu}_\mu$ and
$\tau^+\to\bar{\nu}_\tau\mu^+\nu_\mu$). In order to achieve
a sufficiently fast convergence of the iterative solutions of
the single generalized transport equation of muon neutrinos, one
again has to implement the peculiar $E$ and $X$ dependence as 
implied by the weak $\tau$--decays already into the initial 
zeroth order input $Z^{(0)}(E,X)$.  In this case one needs only
$n=2$ iterations for obtaining a sufficiently accurate result
$Z^{(2)}(E,X)$ for calculating the final secondary $\nu_\mu$
and $\bar{\nu}_\mu$ fluxes.  The $\mu^- +\mu^+$ event rates 
initiated by the secondary neutrinos
are largest obviously at small nadir angles ($\theta<60^{\rm o}$),
with a relative enhancement of at most 40\% over the primary 
$\nu_\mu+\nu_\tau+\tau$ initiated rates for the 
hard initial cosmic fluxes like AGN--M95 and TD--SLBY.  At larger 
nadir angles, 
$\theta$ \raisebox{-0.1cm}{$\stackrel{>}{\sim}$} $60^{\rm o}$, 
the muon rates are dominantly initiated by the primary 
$\nu_\mu+\nu_\tau +\tau$ flux and the secondary $\nu_\mu +
\bar{\nu}_\mu$ flux becomes naturally less relevant.  Thus the 
secondary neutrino flux will enhance the total 
nadir--angle--integrated muon event rates only by less than 10\%.
Nevertheless, it should be possible to observe the effects of
secondary neutrinos with differential $\theta$--dependent
measurements, keeping in mind that the angular resolutions
of the proposed underground neutrino telescopes will reach 
sub--arc--minute precisions. 
\vspace{0.5cm}

\noindent{\underline{\bf Acknowledgements}}

\noindent We are grateful to W.~Rhode for helpful discussions, 
in particular
about lepton ranges, and for providing us with the results of the
$\tau$--ranges of the Chirkin--Rhode Monte Carlo
calculations extended up to $10^{12}$ GeV.  Similarly, we thank 
M.H.~Reno
for sending us her Monte Carlo results for the $\tau$--range for
energies up to $10^{12}$ GeV.  This work has been supported in part
by the   `Bundesministerium f\"ur Bildung und Forschung',
Berlin/Bonn.

\newpage
%Tabelle 1:
%
%
\setlength{\oddsidemargin}{-0.5cm}
\begin{table}[th]
\normalsize
\renewcommand{\arraystretch}{1.2} 
\parbox{17cm}
{\caption[Table 1]{Total nadir--angle--integrated upward--going
$\mu^- +\mu^+$ event rates per year from $(\nu_\tau+\bar{\nu}_\tau)N$
and $(\nu_\mu+\bar{\nu}_\mu)N$ interactions in rock, with the latter
being given in parentheses which are taken from Table 1 of 
\cite{ref5}, for various muon energy thresholds $E_\mu^{\rm min}$
and the appropriate dominant cosmic neutrino fluxes in Fig.~1.
The $\nu_\tau +\bar{\nu}_\tau$ and $\tau^-+\tau^+$ initiated rates
are calculated according to Eqs.~(12) and (14) of Ref.~\cite{ref13},
which are added to the $\nu_\mu+\bar{\nu}_\mu$ initiated rates in
parentheses in order to obtain the final total rates.  A bar signals
that the rates fall below 0.01. This table corrects Table I of
Ref.~\cite{ref13}. \newline}}
\vskip 15pt
\centering 
\begin{tabular}{|l|l||l|l|l|l|l|}
\hline
\raisebox{-1.5ex}[-1.5ex]{Flux} &\raisebox{-1.5ex}[-1.5ex]{Detector}& 
\multicolumn{5}{c|}{Muon-energy threshold E$_{\mu}^{\rm{min}}$/GeV} \\
      &        & $10^3$& $10^4$& $10^5$ & $10^6$& $10^7$\\ \hline\hline
%*****************************************************
       & ANTARES&  16.63 (13.7)&6.28 (5.00)&2.51 (1.98)&1.06 (0.90)&0.34 (0.32)\\
AGN-M95&AMANDA-II& 34.90 (29.1)&10.76 (8.62)&3.78 (2.98)&1.58 (1.34)&0.49 (0.46)\\      
       & IceCube&  170.24 (143)&41.72 (33.7)&14.22 (11.2)&5.93 (5.04)&1.83 (1.74)\\
\hline
%******************************************************
      & ANTARES&  0.75 (0.60)&0.39 (0.32)&0.10 (0.08)&0.01 (0.01)&--- \\ 
GRB-WB&AMANDA-II& 1.39 (1.10)&0.68 (0.56)&0.15 (0.13)&0.02 (0.02)&--- \\  
      & IceCube&  5.55 (4.35)&2.59 (2.13)&0.57 (0.49)&0.07 (0.06)&--- \\
\hline
%******************************************************
      & ANTARES&  0.84 (0.62)&0.59 (0.45)&0.33 (0.26)&0.14 (0.12)&0.05 (0.05)\\ 
TD-SLBY&AMANDA-II&1.33 (0.97)&0.91 (0.68)&0.49 (0.39)&0.21 (0.18)&0.07 (0.07)\\  
      & IceCube&  5.11 (3.70)&3.42 (2.57)&1.84 (1.47)&0.78 (0.68)&0.26 (0.25)\\
\hline
%******************************************************
\end{tabular}
\end{table}

\newpage
%%Tabelle 2:
%
\setlength{\oddsidemargin}{-0.5cm}
\begin{table}[th]
\normalsize
\renewcommand{\arraystretch}{1.2} 
\parbox{17cm}
{\caption[Table 2]{Nadir--angle--integrated upward--going 
$\mu^- +\mu^+$ event rates per year for muons with energy above 
$E_\mu^{\rm min}= 10^4$ GeV.  The events produced by the primary
$\nu_\mu +\bar{\nu}_\mu$, $\nu_\tau+\bar{\nu}_\tau$ and 
$\tau^- +\tau^+$ fluxes are given in parentheses, which are obtained
from appropriately integrating the relevant dashed--dotted curves
in Fig.~20.  Notice that the total ($0^{\rm o}\leq\theta\leq 90^{\rm o}$)
event rates in brackets in the last column agree of course with
the final total rates in Table 1.  Adding to these conventional
primary rates the ones induced by the secondary 
$\nu_\mu+\bar{\nu}_\mu$ fluxes, originating from 
$\tau^+\to\nu_\mu$ and $\tau^-\to \bar{\nu}_\mu$, 
one obtains the final results
shown.  These additional secondary $\nu_\mu +\bar{\nu}_\mu$ 
contributions are calculated according to (39) and correspond to 
integrating appropriately the relevant solid curves in 
Fig.~20. \newline}}
\vskip 15pt
\centering 
\begin{tabular}{|l|l||l|l|l|l|}
\hline
\raisebox{-1.5ex}[-1.5ex]{Flux} &\raisebox{-1.5ex}[-1.5ex]{Detector}& 
\multicolumn{4}{c|}{Number of events} \\
      &        & $0^\circ\le\theta\le 30^\circ$& $30^\circ\le\theta\le 60^\circ$& $60^\circ\le\theta \le90^\circ$ & Total\\ \hline\hline
%*****************************************************
       & ANTARES&  0.19 (0.18)&1.13 (1.05)&5.25 (5.04) &6.58 (6.28)\\ 
AGN-M95&AMANDA-II& 0.40 (0.38)&2.15 (2.03)&8.66 (8.34) &11.22 (10.76)\\
       & IceCube&  1.62 (1.54)&8.54 (8.06)&33.35 (32.13)&43.50 (41.72)\\ 
\hline
%******************************************************
      & ANTARES&   0.02 (0.01)&0.10 (0.09)&0.28 (0.28)&0.39 (0.39)\\ 
GRB-WB&AMANDA-II&  0.03 (0.03)&0.18 (0.17)&0.48 (0.48)&0.69 (0.68)\\ 
      & IceCube&   0.12 (0.11)&0.69 (0.67)&1.84 (1.82)&2.65 (2.59)\\
\hline
%******************************************************
      & ANTARES&    0.01 (0.01)&0.07 (0.06)&0.54 (0.52)&0.62 (0.59)\\
TD-SLBY&AMANDA-II&  0.01 (0.01)&0.12 (0.10)&0.83 (0.79)&0.96 (0.91)\\
      & IceCube&    0.05 (0.04)&0.44 (0.38)&3.14 (3.00)&3.63 (3.42)\\
\hline
%******************************************************
\end{tabular}
\end{table}
\newpage

%Tabelle 3:
%
\setlength{\oddsidemargin}{-0.5cm}
\begin{table}[th]
\normalsize
\renewcommand{\arraystretch}{1.2} 
\parbox{17cm}
{\caption[Table 3]{As in Table 2 but for $E_\mu^{\rm min} =10^5$ GeV.
A bar signals that the rates fall below $0.01$.\newline}}
\vskip 15pt
\centering 
\begin{tabular}{|l|l||l|l|l|l|}
\hline
\raisebox{-1.5ex}[-1.5ex]{Flux} &\raisebox{-1.5ex}[-1.5ex]{Detector}& 
\multicolumn{4}{c|}{Number of events} \\
      &        & $0^\circ\le\theta\le 30^\circ$& $30^\circ\le\theta\le 60^\circ$& $60^\circ\le\theta \le90^\circ$ & Total\\ \hline\hline
%*****************************************************
       & ANTARES&  0.01 (0.01)&0.20 (0.17)&2.46 (2.33)&2.67 (2.51)\\ 
AGN-M95&AMANDA-II& 0.02 (0.02)&0.31 (0.26)&3.68 (3.50)&4.01 (3.78)\\ 
       & IceCube&  0.08 (0.06)&1.16 (0.98)&13.89 (13.20)&15.11 (14.22)\\
\hline
%******************************************************
      & ANTARES&   ---        & 0.02 (0.02)&0.08 (0.08)&0.10 (0.10)\\ 
GRB-WB&AMANDA-II&  ---        & 0.03 (0.03)&0.12 (0.12)& 0.15 (0.15)\\   
      & IceCube&   0.01 (0.01)& 0.11 (0.10)&0.47 (0.46)& 0.58 (0.57)\\  
\hline
%******************************************************
      & ANTARES&   ---        & 0.03 (0.02)&0.32 (0.30)& 0.34 (0.33)\\ 
TD-SLBY&AMANDA-II& ---        & 0.04 (0.03)&0.48 (0.45)& 0.52 (0.49)\\ 
      & IceCube&   0.01 (0.01)& 0.14 (0.12)&1.79 (1.71)& 1.94 (1.84)\\ 
\hline
%******************************************************
\end{tabular}
\end{table}

\clearpage
\begin{figure}
\epsfig{file=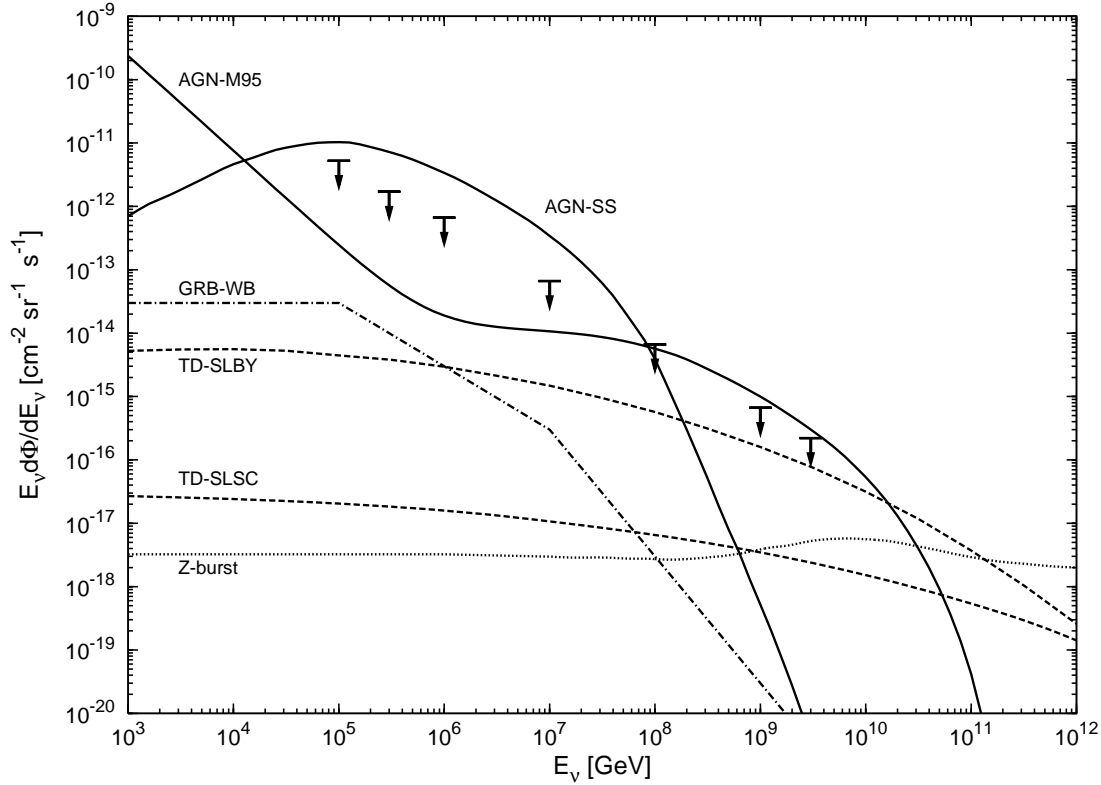,angle=270,width=15cm}
\caption{Representative differential fluxes of muon neutrinos 
  $(\nu_\mu+\bar{\nu}_\mu)$ at the production site from active galactic
  nuclei (AGN--SS \cite{ref22} and AGN--M95 \cite{ref23}), gamma ray
  bursts (GRW-WB \cite{ref24}), toplogical defects (TD--SLBY \cite{ref25}
  and TD--SLSC \cite{ref26}) and $Z$--bursts \cite{ref27}. Due to naive
  channel counting in pion production and decay at the production site
  ($\nu_e:\nu_\mu:\nu_\tau = 1:2:0$) and maximal mixing, 
  $\nu_e:\nu_\mu:\nu_\tau = 1:1:1$, these fluxes are divided equally
  between $e$--, $\mu$-- and $\tau$--neutrinos when they reach the 
  Earth's surface (i.e.\ will be divided by a factor of 2).  The
  diffuse neutrino flux upper limit of AMANDA \cite{ref20,ref21} 
  are shown by the bars with arrows.}
\end{figure}

\clearpage
\begin{figure}
\epsfig{file=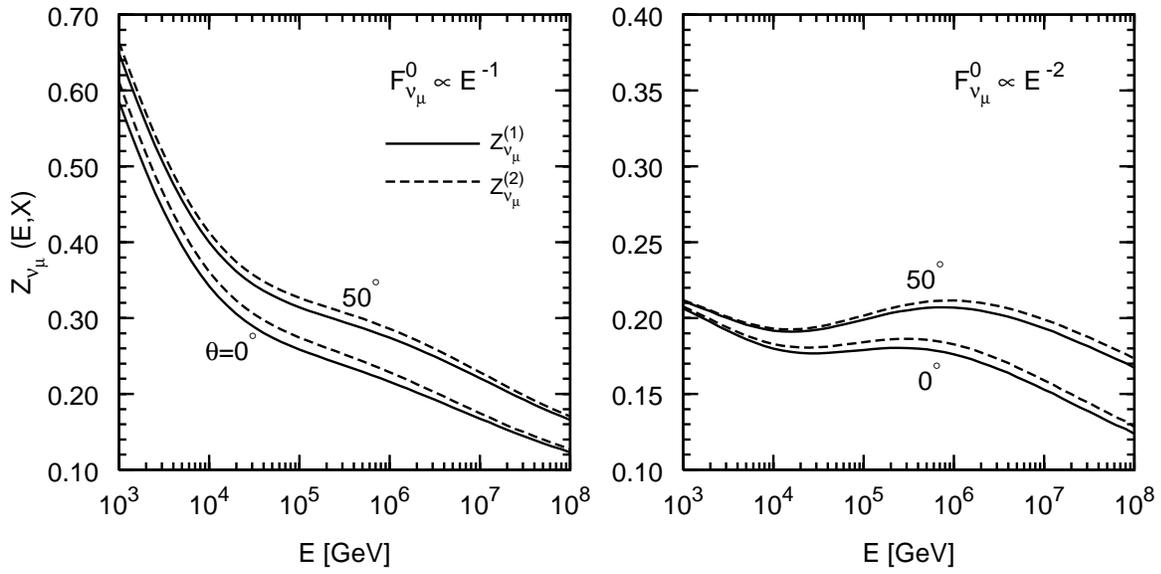, width=\textwidth}
\caption{The $Z_\nu$--factors for $\nu_\mu$ neutrinos, as iteratively
  calculated according to (6), for the generic initial neutrino fluxes
  in (10) and (11) divided by 2.  The result for the first order
  iteration $Z_{\nu_\mu}^{(1)}$ is given in (8).  For nadir angles
  $\theta >50^{\rm o}$, the second order iterative result 
  $Z_{\nu_\mu}^{(2)}$ becomes almost indistinguishable from
  $Z_{\nu_\mu}^{(1)}$.}
\end{figure}

\clearpage
\begin{figure}
\epsfig{file=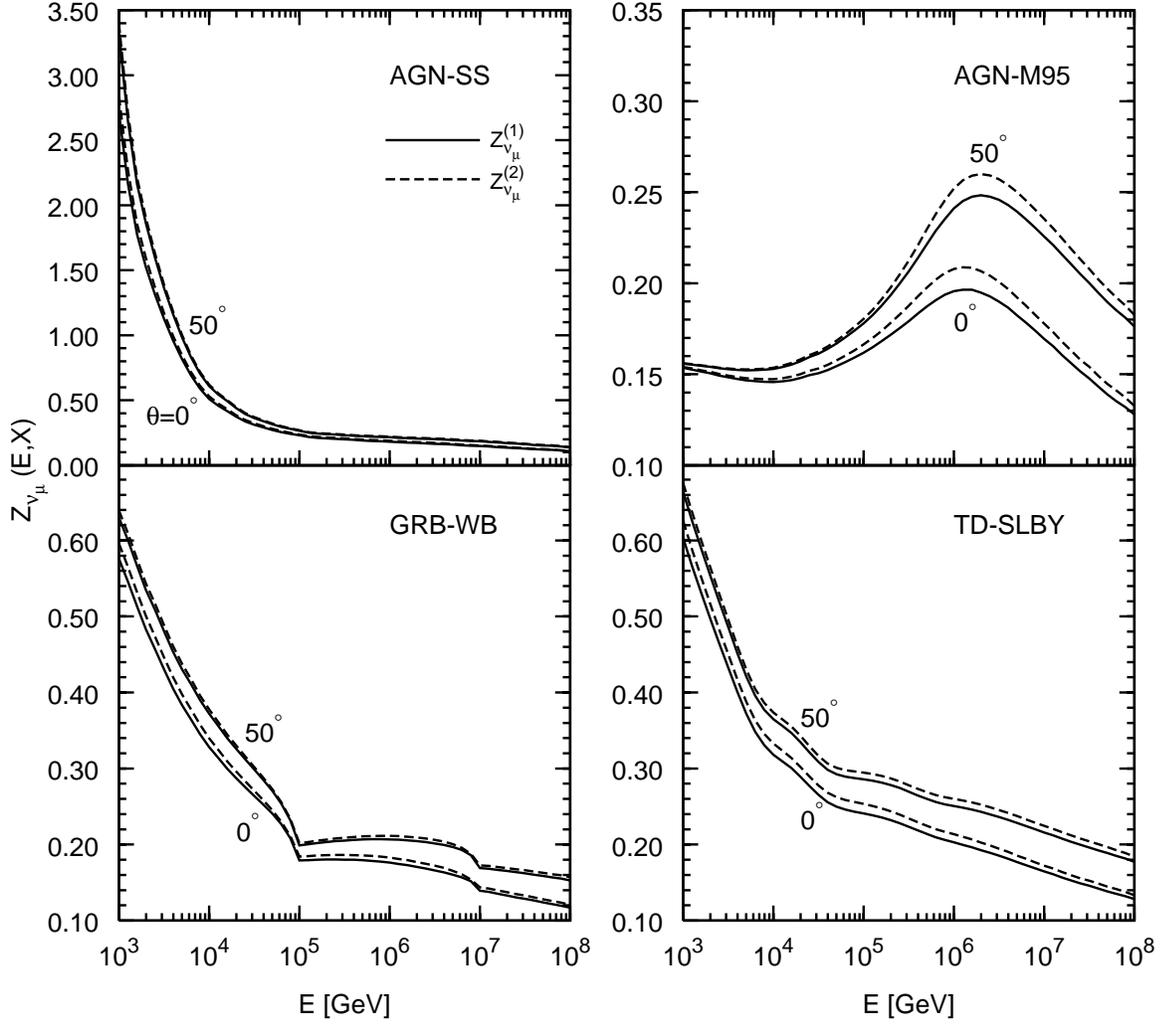, width=\textwidth}
\caption{As in Fig.~2 but for some typical initial cosmic fluxes shown
  in Fig.~1.  The (small) TD--SLSC and $Z$--burst fluxes in Fig.~1
  result in a similar iterative convergence as the $F_{\nu_\mu}^0
  \sim E^{-1}$ flux in Fig.~2.}
\end{figure}

\clearpage
\begin{figure}
\epsfig{file=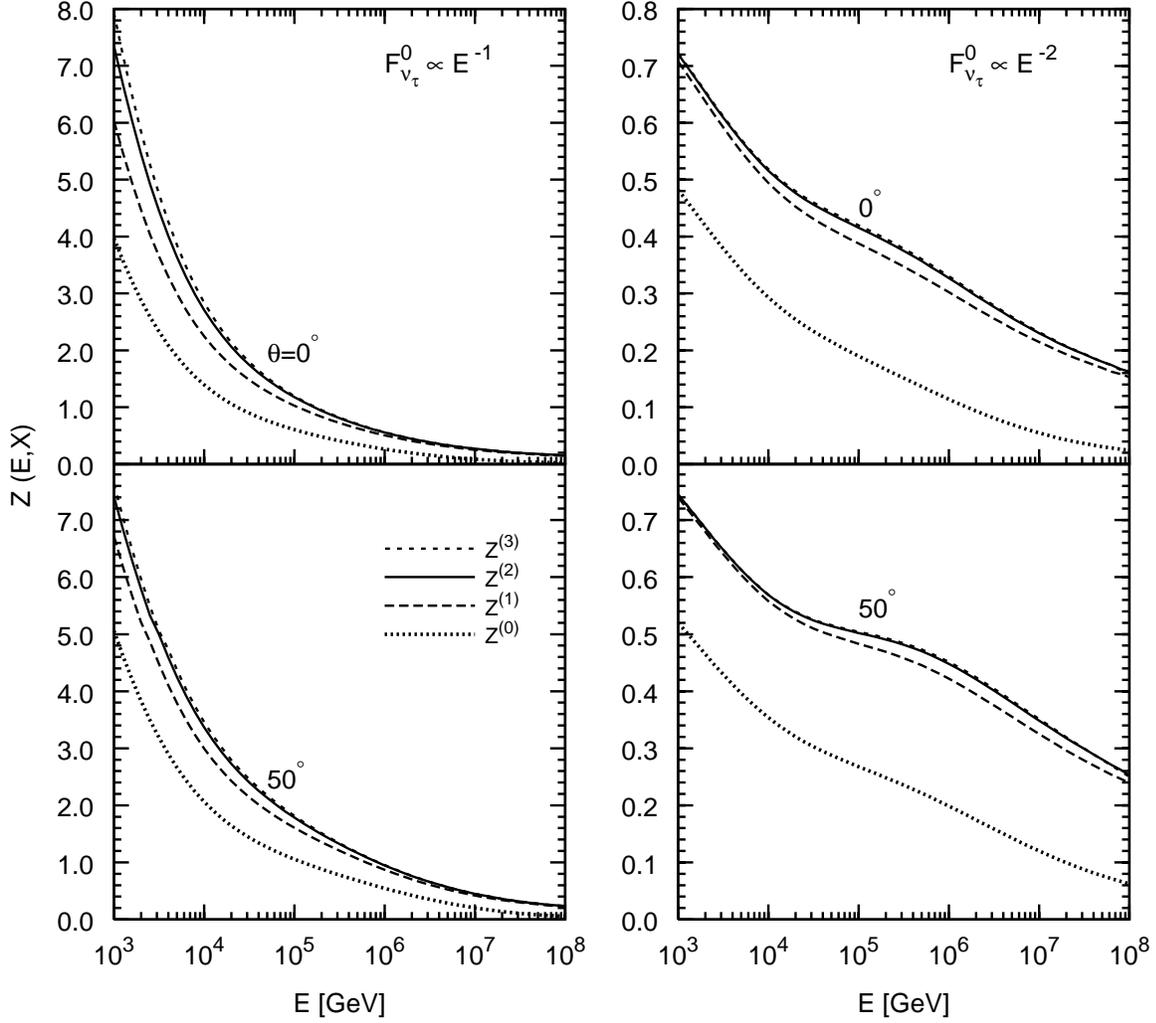,width=\textwidth}
\caption{The $Z$--factors for $\nu_\tau$ neutrinos, 
  $Z^{(n)}=Z_{\nu_\tau}^{(n)}+ Z_{\tau^-}^{(n)}$, 
  as iteratively calculated for $n=1,\,2,\, 3$ iterations using the
  input $Z^{(0)}$ of (21) which is shown by the dotted curves.
  The generic initial fluxes are taken from (10) and (11) divided
  by 2.} 
\end{figure}

\clearpage
\begin{figure}
\epsfig{file=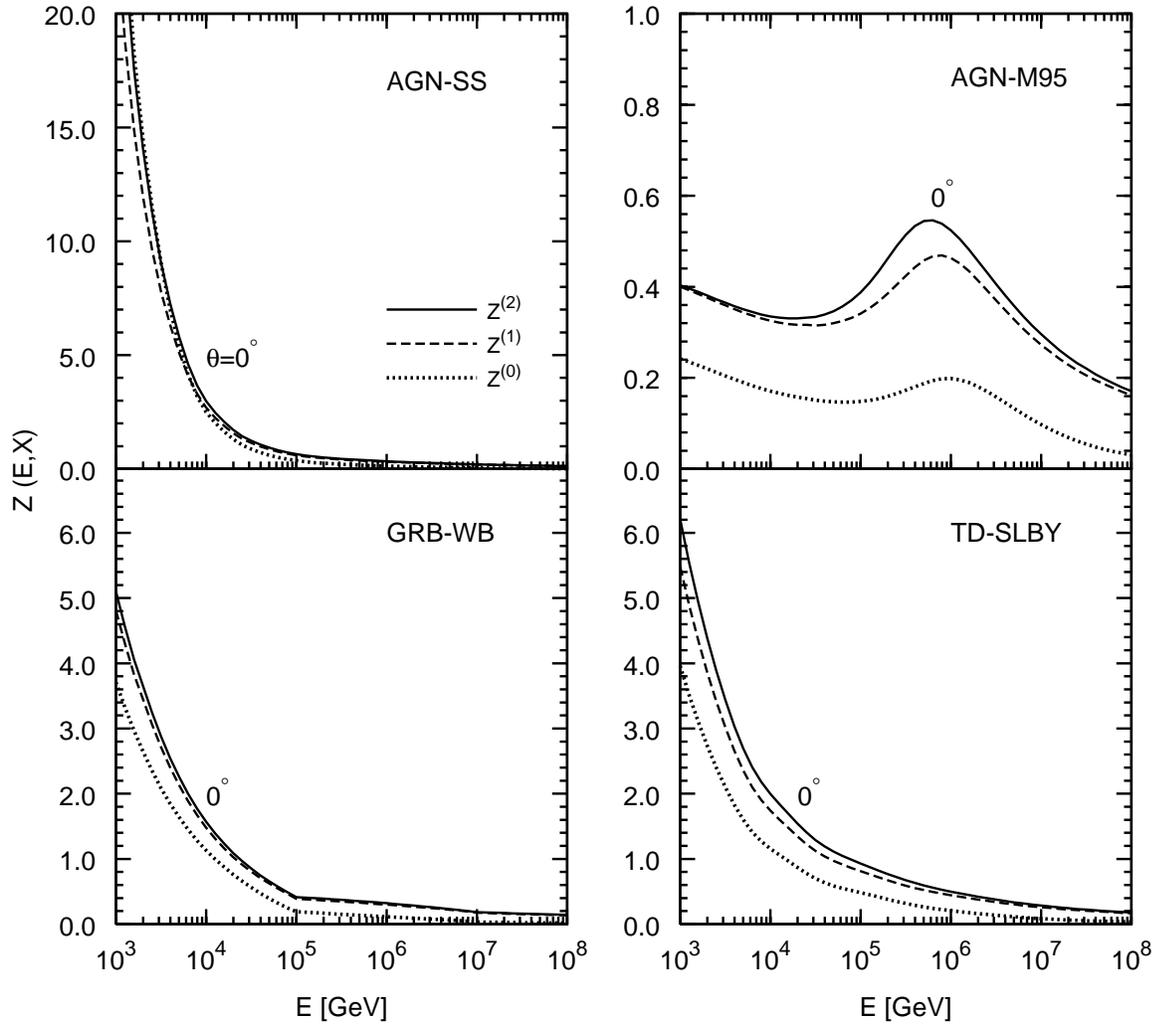,width=\textwidth}
\caption{As in Fig.~4 but only for the relevant $n=1,\, 2$ iterations
for $\theta=0^{\rm o}$ and for the dominant initial cosmic fluxes
in Fig.~1.}  
\end{figure}

\clearpage
\begin{figure}
\epsfig{file=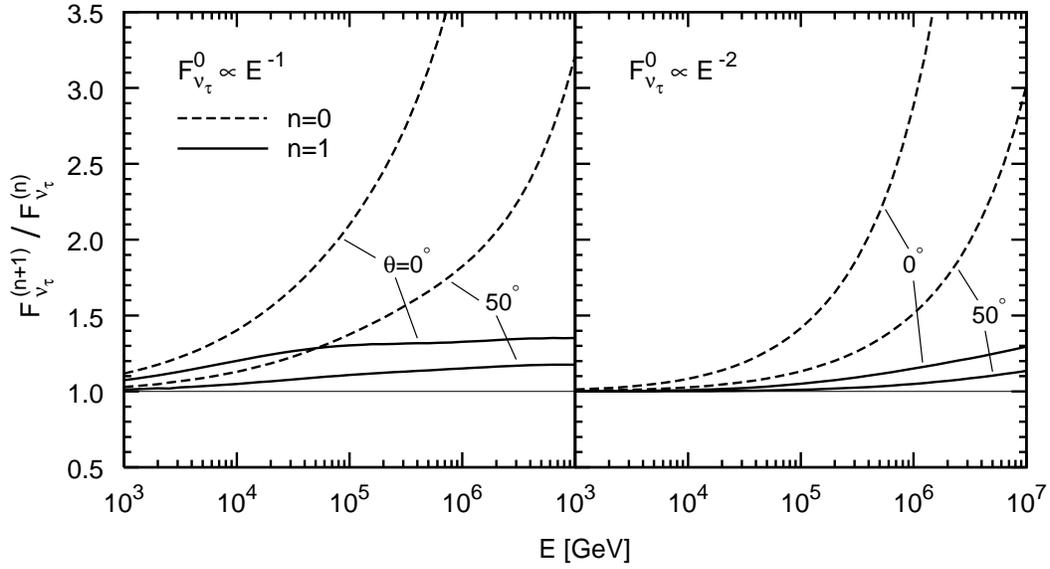,width=\textwidth}
\caption{The ratios of the $\nu_\tau$ fluxes 
$F_{\nu_\tau}^{(n+1)}/F_{\nu_\tau}^{(n)}$ according to two consecutive
iterations.  The fluxes are calculated according to (24) using the 
appropriate $Z^{(n)}$ factors shown in Fig.~4 for the
two generic $E^{-1}$ and $E^{-2}$ initial fluxes.}
\end{figure}

\clearpage
\begin{figure}
\epsfig{file=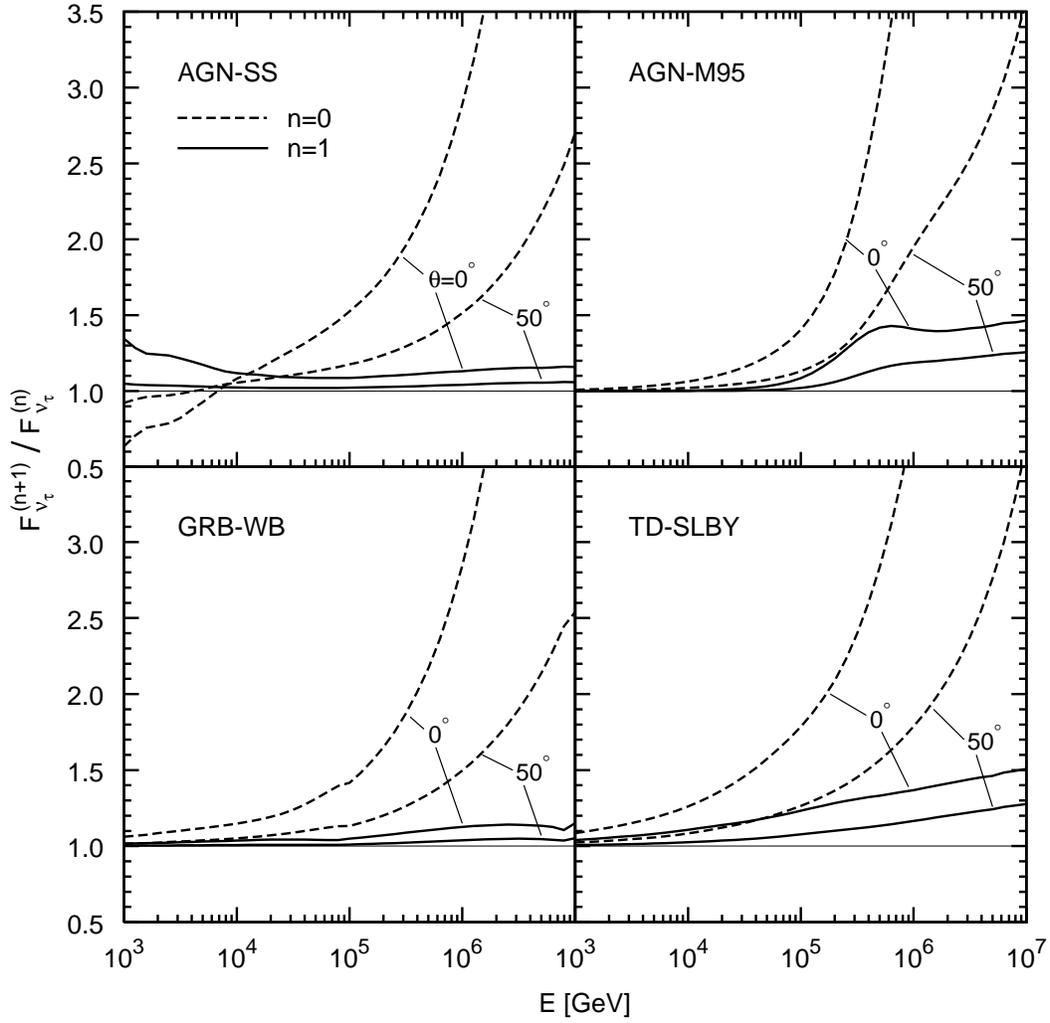,width=\textwidth}
\caption{As in Fig.~6 but with the appropriate $Z^{(n)}$ factors shown
in Fig.~5 as obtained for the dominant initial cosmic fluxes in Fig.~1.}
\end{figure}

\clearpage
\begin{figure}
\epsfig{file=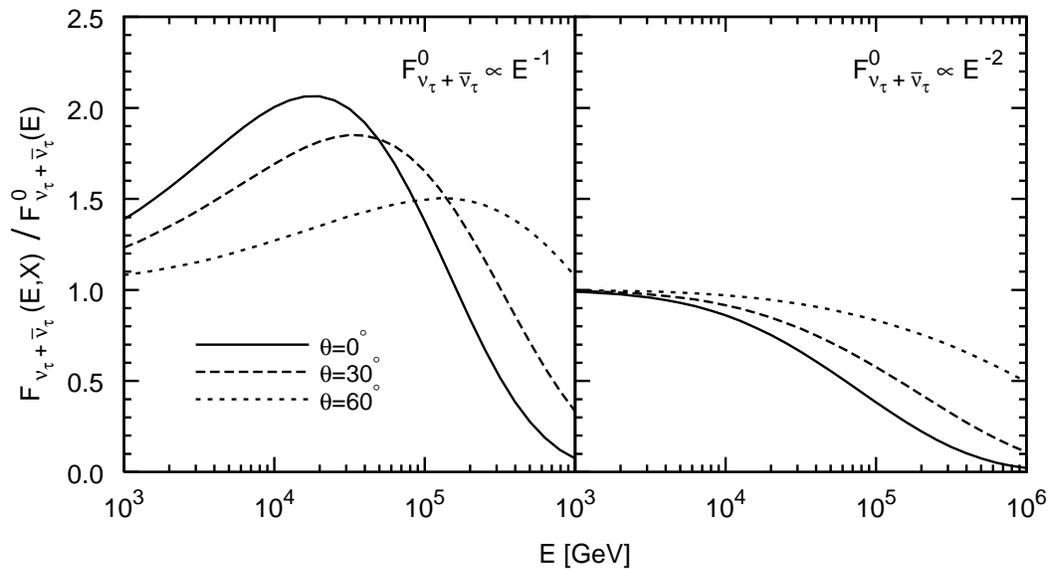,width=\textwidth}
\caption{The final total $\nu_\tau+\bar{\nu}_\tau$ fluxes, calculated 
according to (24) using the appropriate iterative $n=2$ results 
$Z^{(2)}$ for $\theta =0^{\rm o}$, $30^{\rm o}$ and $60^{\rm o}$
($Z^{(2)}$ for $\theta=0^{\rm o}$ is shown in Fig.~4). The generic
initial fluxes $F_{\nu_\tau+\bar{\nu}_\tau}^0(E)$ are given in (10)
and (11).} 
\end{figure}

\clearpage
\begin{figure}
\epsfig{file=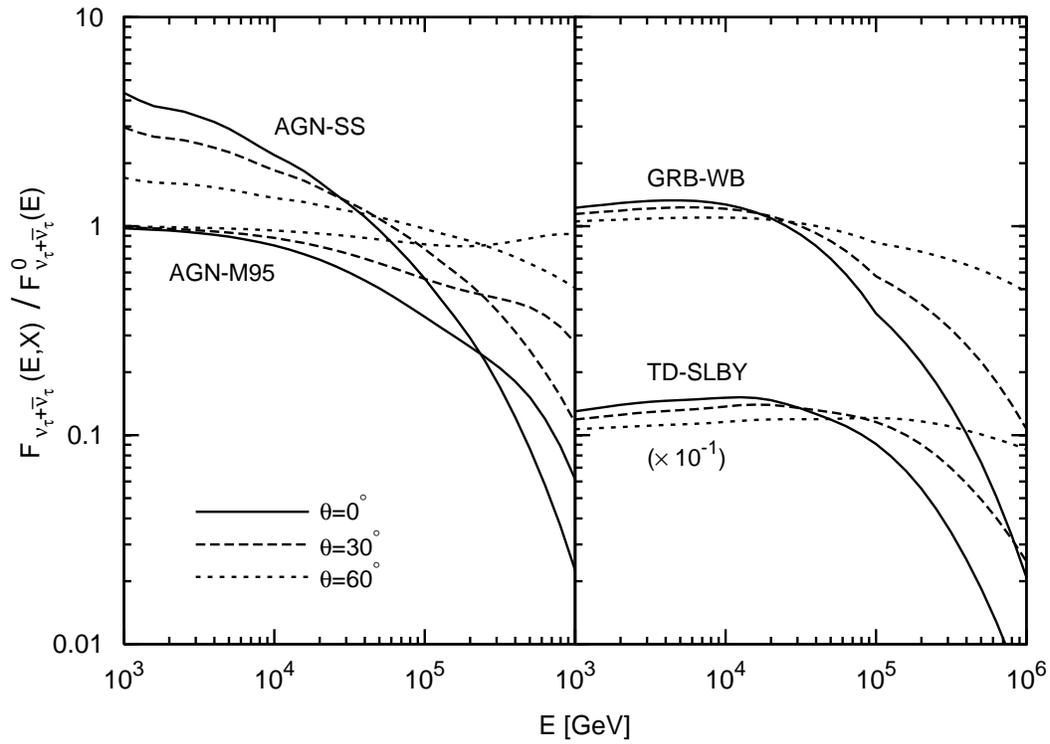, width=\textwidth}
\caption{As in Fig.~8 but for the dominant initial cosmic fluxes
$F_{\nu_\tau +\bar{\nu}_\tau}^0(E)$ in Fig.~1. For $\theta=0^{\rm o}$
the relevant $Z^{(2)}$ factors are shown in Fig.~5. The TD--SLBY 
results are multiplied by $10^{-1}$ as indicated.}
\end{figure}

\clearpage
\begin{figure}
\centering
\epsfig{file=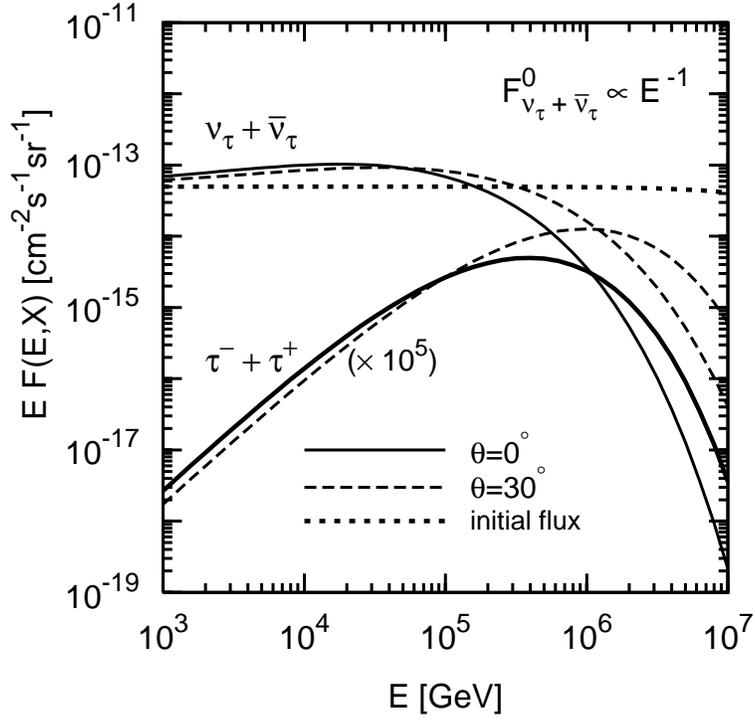,width=12cm}
\caption{Attenuated and regenerated $\nu_\tau+\bar{\nu}_\tau$ and
$\tau^-+\tau^+$ fluxes calculated according to (24) and (25), 
respectively, using the sufficiently accurate second iterative
solution $Z^{(2)}(E,X)$ for nadir angles $\theta=0^{\rm o}$ and
$30^{\rm o}$, and the generic initial 
$F_{\nu_\tau +\bar{\nu}_\tau}^0(E)$ flux in (10) which is shown by
the dotted curve.  The results for the $\tau^- +\tau^+$ fluxes are
multiplied by $10^5$ as indicated.}
\end{figure}

\clearpage
\begin{figure}
\epsfig{file=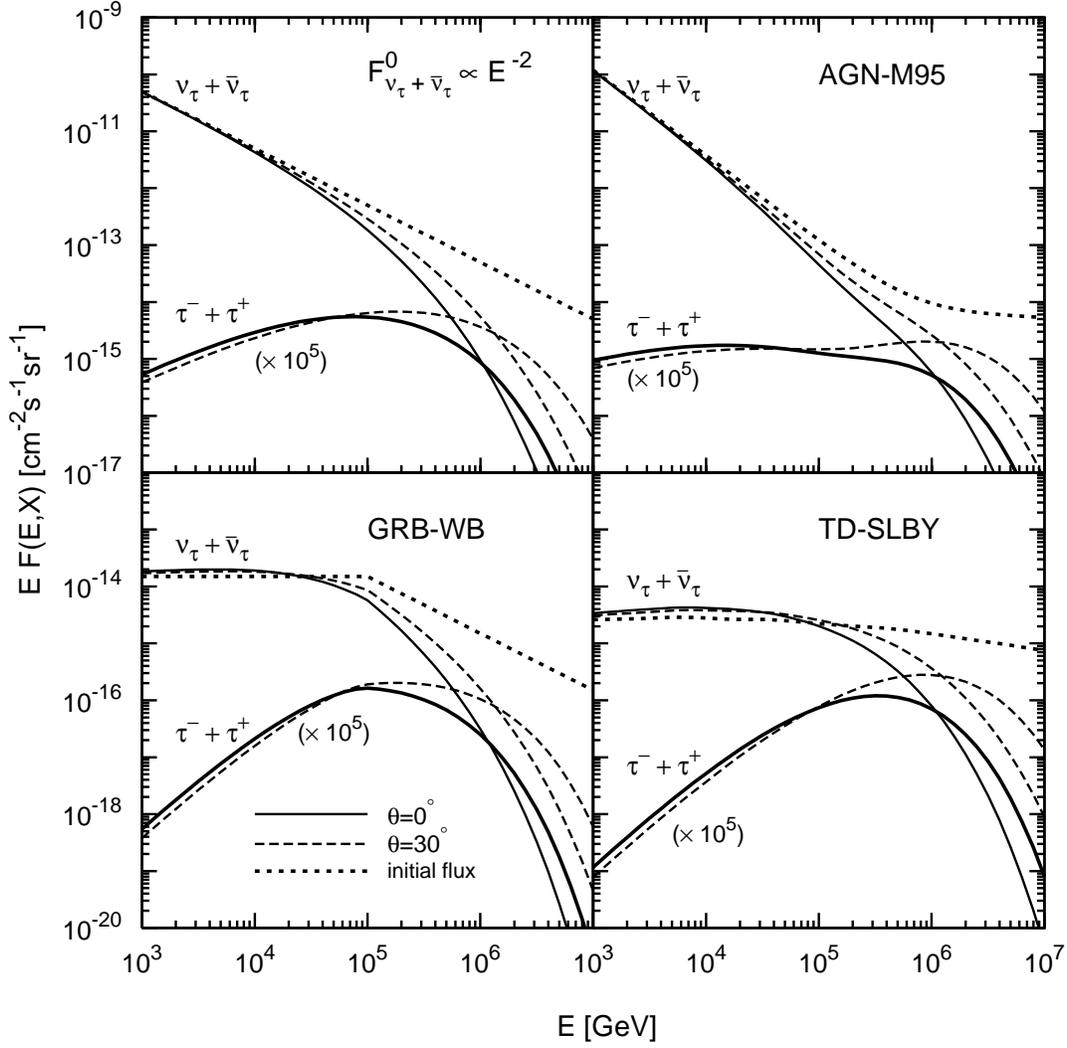, width=\textwidth}
\caption{As in Fig.~10 but for the generic $E^{-2}$ flux in (11) and the
relevant dominant initial fluxes $F_{\nu_\tau+\bar{\nu}_\tau}^0 =
\frac{1}{2}\, d\Phi/dE_\nu$ in Fig.~1 which are shown by the dotted 
curves.}
\end{figure}

\clearpage
\begin{figure}
\epsfig{file=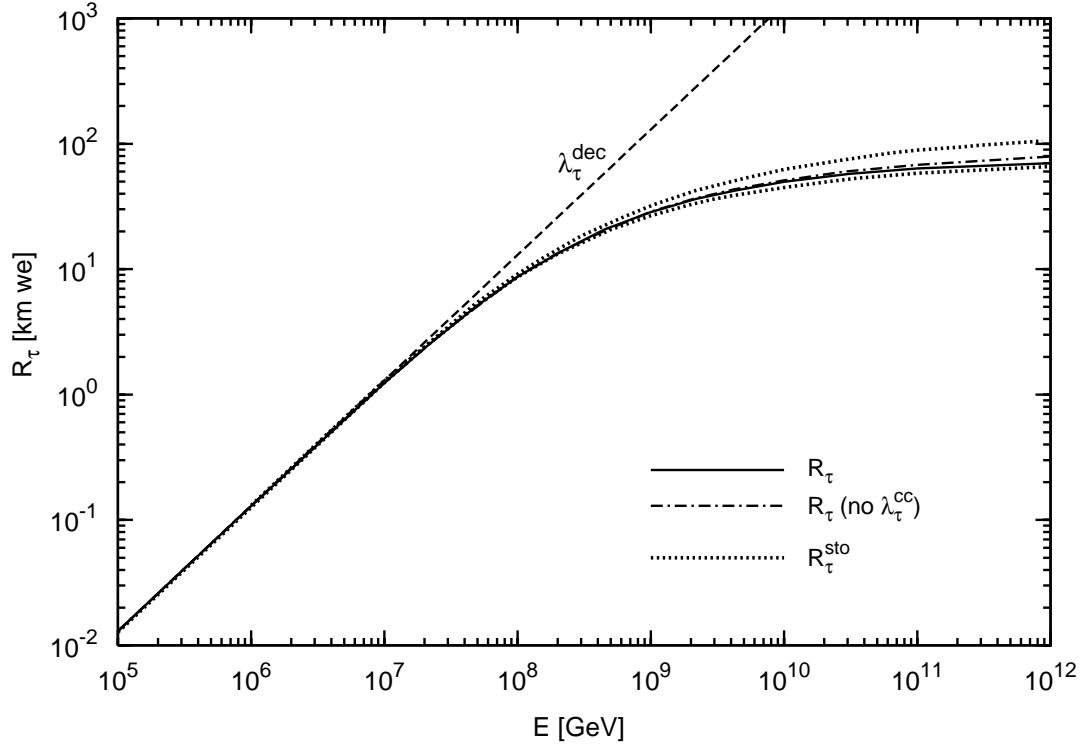,width=\textwidth}
\caption{The $\tau$--decay length 
$\lambda_\tau^{\rm dec}=(E/m_\tau)c\tau_\tau\rho$, shown by the dashed
curve, and the $\tau$--ranges in standard rock ($\rho=2.65$ g/cm$^3$)
for an incident $\tau$--energy $E$ and final $\tau$--energy larger
than $E^{\rm min} =50$ GeV.  $R_\tau$ is calculated according to (27)
and (28); omitting $\lambda_\tau^{\rm CC}$ in (27) results in 
$R_\tau$(no $\lambda_\tau^{\rm CC}$).  The stochastic Monte Carlo
evaluations of $R_\tau^{\rm sto}$ are shown by the upper dotted curve
\cite{ref38} and the lower one according to \cite{ref32,ref35} which
are based on the ALLM97 parametrization of structure functions for
calculating the photonuclear cross section.  Using a different 
parametrization (Bugaev--Schlepin) for extrapolating the latter
cross section to ultrahigh energies, results in a $\tau$--range which
is about 25\% smaller than the upper dotted curve at $E=10^{12}$ GeV
\cite{ref38}.}
\end{figure}

\clearpage
\begin{figure}
\epsfig{file=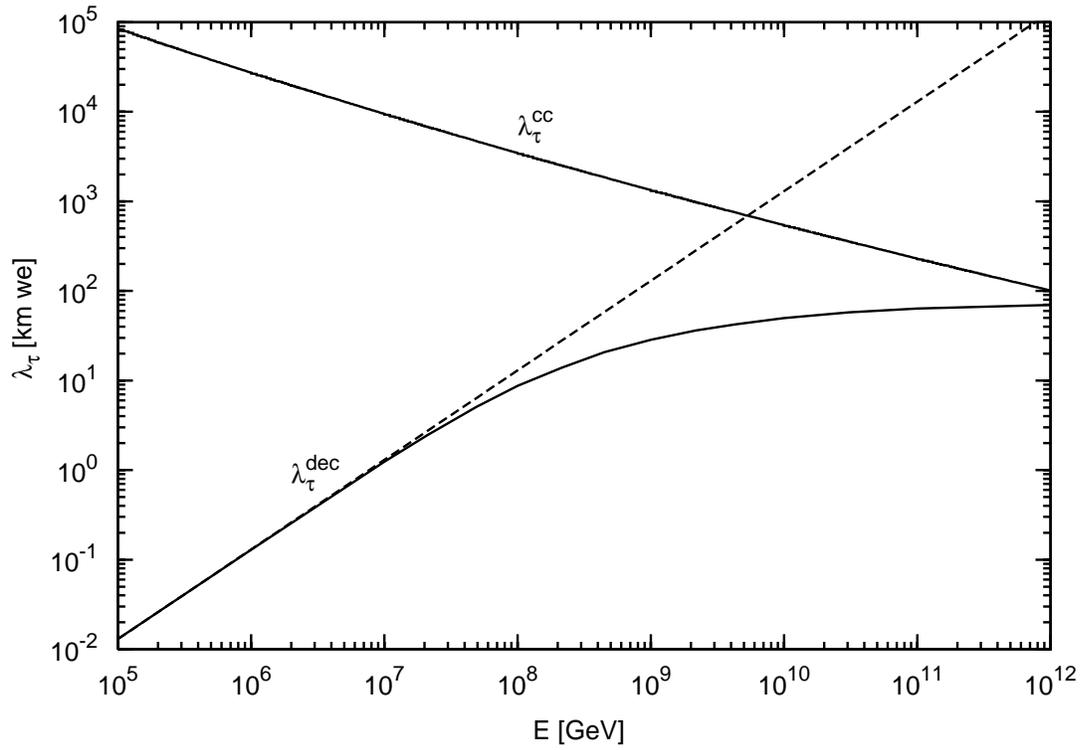,width=\textwidth}
\caption{The CC $\tau$--interaction length $\lambda_\tau^{\rm CC}$
(upper solid curve) and the $\tau$--decay length neglecting the
energy--loss (dashed curve) and including the energy--loss in 
standard rock (lower solid line). The latter two curves correspond
to the ones shown in Fig.~12.}
\end{figure}

\clearpage
\begin{figure}
\epsfig{file=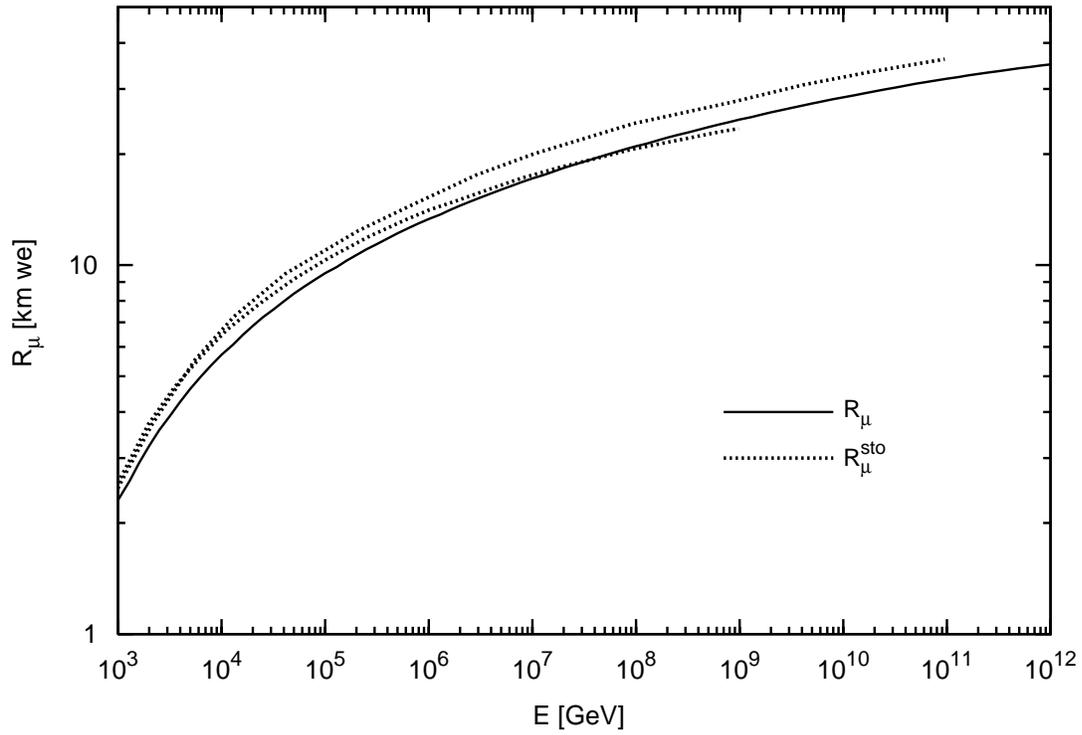,width=\textwidth}
\caption{The $\mu$--ranges in standard rock ($\rho =2.65$ g/cm$^3$)
for an incident muon energy $E$ and the final $\mu$--energy larger
than $E^{\min}=1$ GeV.  $R_\mu$ is calculated according to (27) and
(28) as discussed in the text.  The stochastic Monte Carlo evaluations
of $R_\mu^{\rm sto}$ are taken from $\cite{ref32,ref35}$ (lower
dotted curve) and from \cite{ref38} (upper dotted curve).}
\end{figure}

\clearpage
\begin{figure}
\centering
\epsfig{file=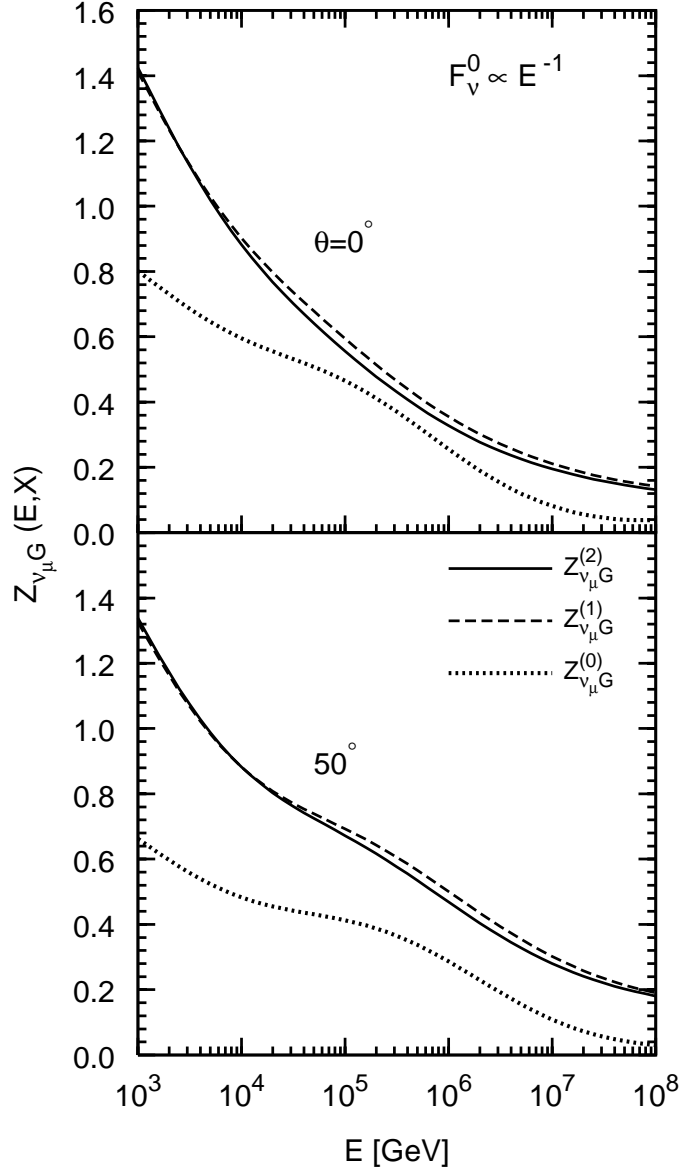,width=11.0cm}
\caption{The $Z$--factors for primary and secondary $\nu_\mu$ neutrinos,
$Z_{\nu_\mu G}^{(n)}=Z_{\nu_\mu}^{(n)}+Z_G^{(n)}$, as defined in
(34), (35), (36) and calculated iteratively for $n=1,\, 2$ according to
(37) using the input $Z_{\nu_\mu G}^{(0)}$ of (38) which is shown by
the dotted curves.  The generic initial $E^{-1}$ flux is taken from
(10) divided by 2.}
\end{figure}

\clearpage
\begin{figure}
\epsfig{file=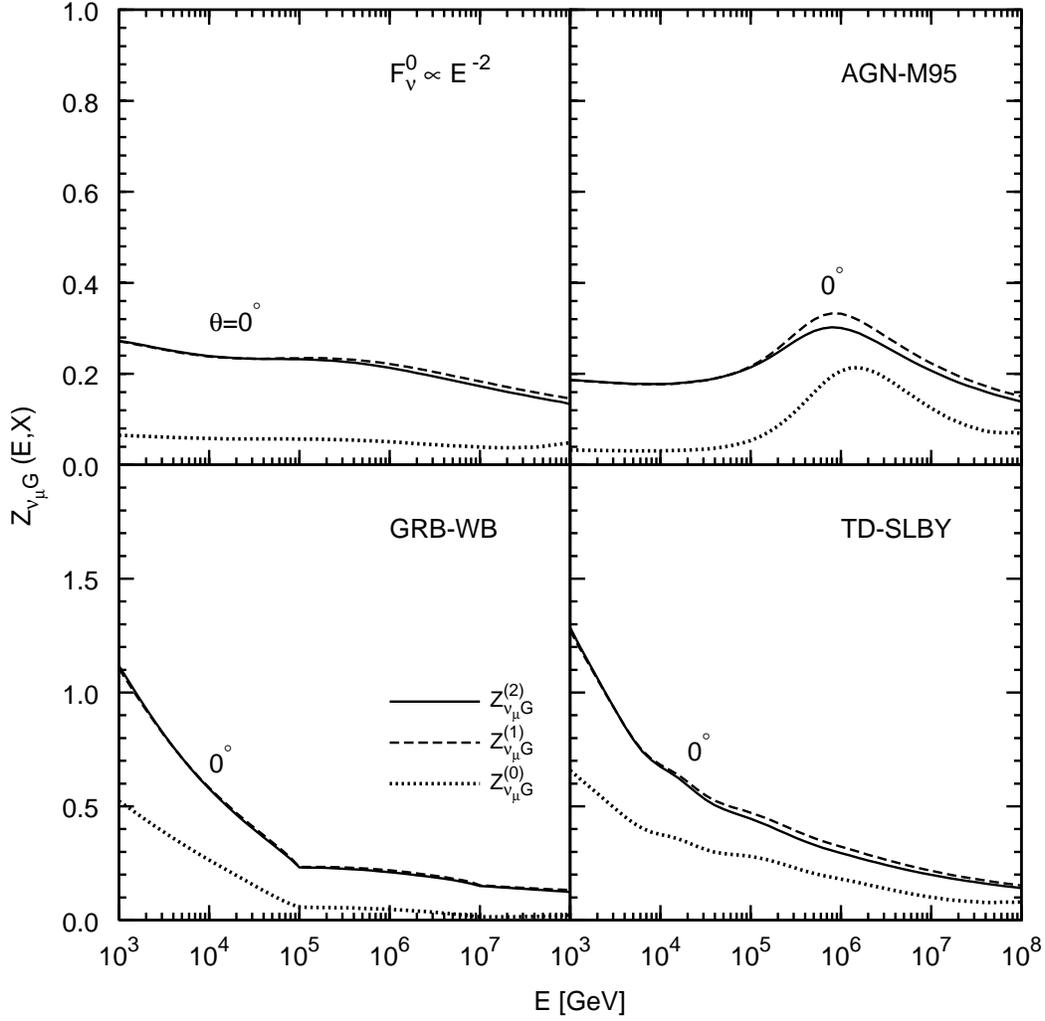,width=\textwidth}
\caption{As in Fig.~15 but only for $\theta=0^{\rm o}$ and for the
generic initial $E^{-2}$ flux in (11), divided by 2, and the three
dominant initial cosmic fluxes in Fig.~1.}
\end{figure}

\clearpage
\begin{figure}
\epsfig{file=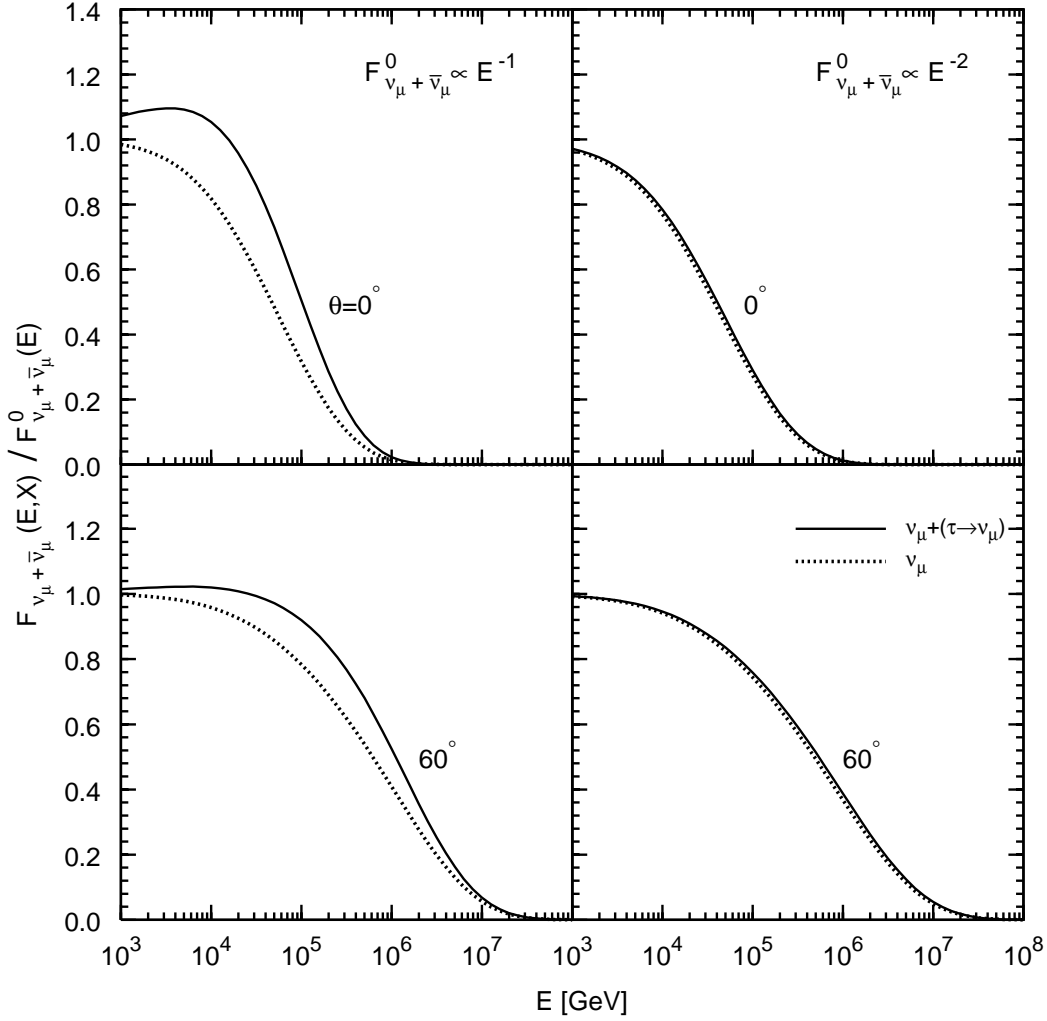,width=\textwidth}
\caption{The final total $\nu_\mu +\bar{\nu}_\mu$ fluxes for primary
and secondary muon neutrinos, calculated according to (33) using the
appropriate iterative $n=2$ results for $Z_{\nu_\mu G}^{(2)}$ for 
$\theta= 0^{\rm o}$ and $60^{\rm o}$.  The generic initial fluxes 
$F_{\nu_\mu +\bar{\nu}_\mu}^0(E)$ are given in (10) and (11).  The
usual primary $\nu_\mu +\bar{\nu}_\mu$ fluxes are for brevity denoted by
$\nu_\mu$ (dotted curves).  Similarly the secondary neutrino
contributions due $\tau^+ \to\nu_\mu$ and $\tau^- \to\bar{\nu}_\mu$ are
denoted by $\tau\to\nu_\mu$.}
\end{figure}

\clearpage
\begin{figure}
\epsfig{file=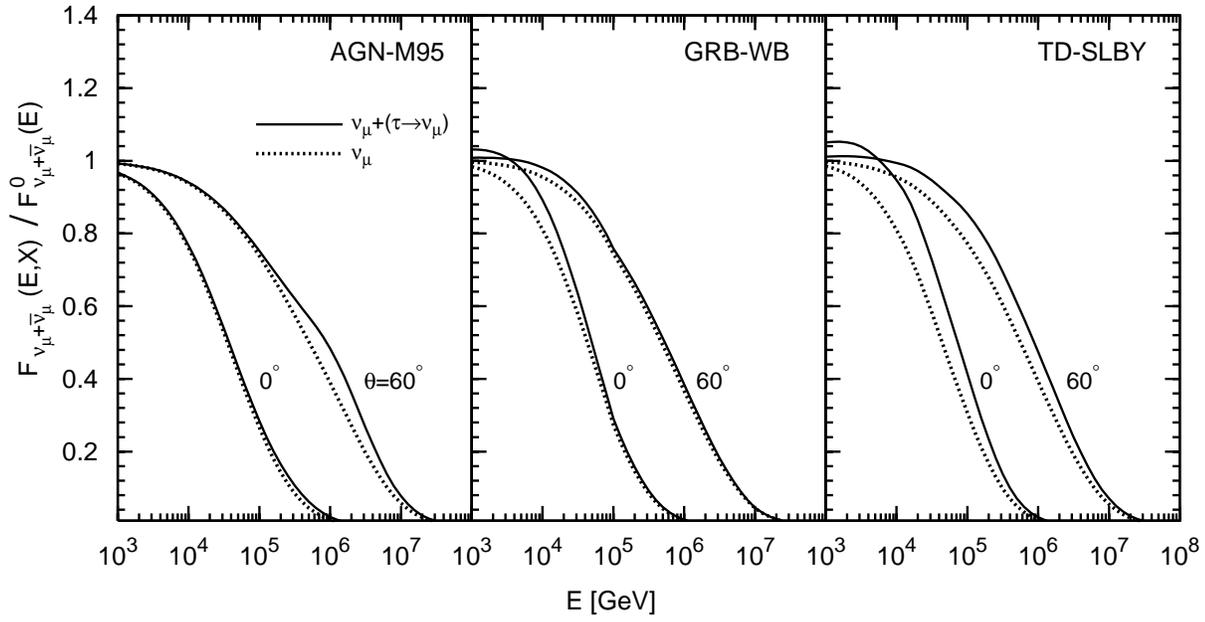,width=\textwidth}
\caption{As in Fig.~17 but for the three dominant initial cosmic
fluxes $F_{\nu_\mu +\bar{\nu}_\mu}^0(E)$ in Fig.~1.}
\end{figure}

\clearpage
\begin{figure}
\epsfig{file=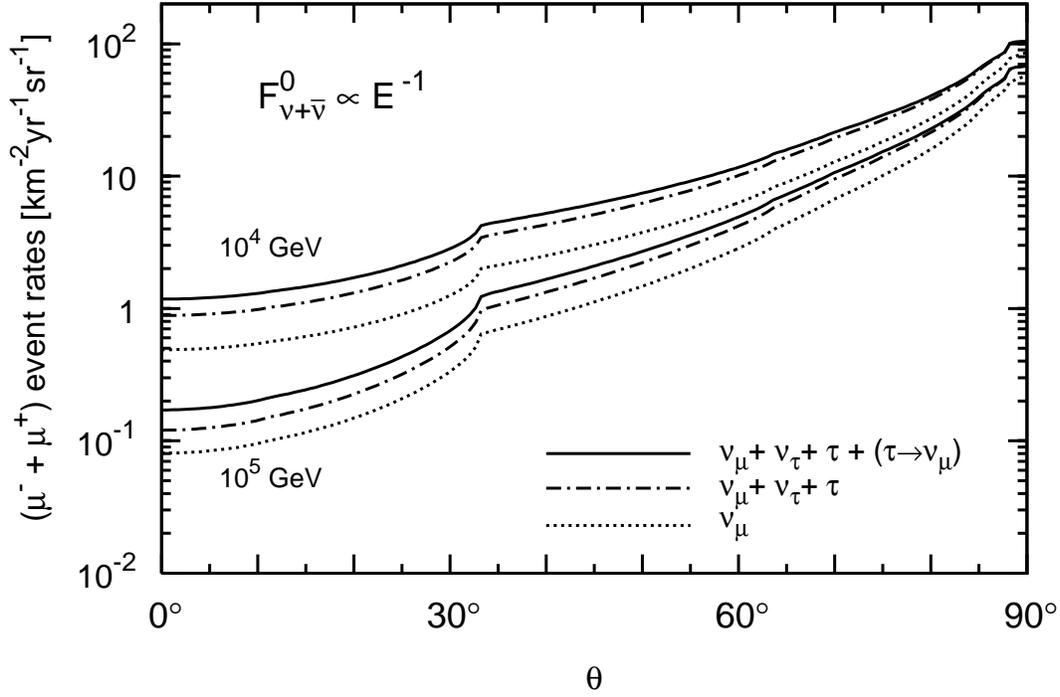,width=\textwidth}
\caption{The $\mu^- +\mu^+$ event rates for muons with energy 
above $10^4$ and $10^5$ GeV originating from (anti)neutrinos 
with $E_\nu\leq 10^8$ GeV for the generic initial $E^{-1}$ flux 
in (10). The muon events initiated by the primary 
($\nu_\mu+\bar{\nu}_\mu$) and $(\nu_\mu +\bar{\nu}_\mu)+
(\nu_\tau+\bar{\nu}_\tau)+(\tau^-+\tau^+)$ fluxes are for 
simplicity denoted by $\nu_\mu$ and $\nu_\mu +\nu_\tau +\tau$, 
respectively (dotted and dashed--dotted curves).  
The additional events initiated by the secondary muon neutrinos 
arising from $\tau^-\to \bar{\nu}_\mu$,
$\tau^+\to\nu_\mu$ are for brevity denoted by ($\tau\to\nu_\mu$).} 
\end{figure}

\clearpage
\begin{figure}
\epsfig{file=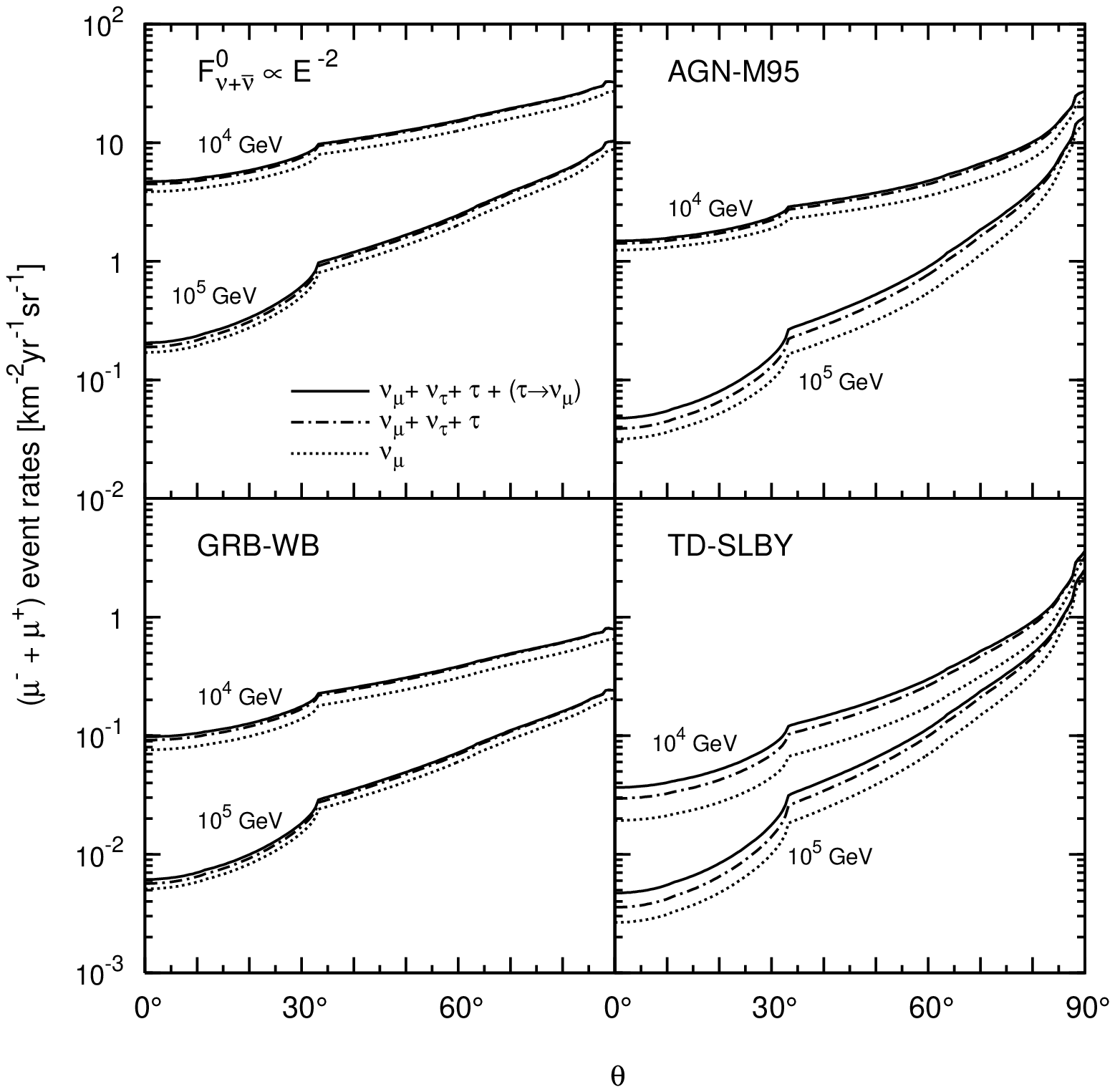,width=\textwidth}
\caption{As in Fig.~19 but for the generic initial $E^{-2}$ flux 
in (11) and for the three dominant initial cosmic fluxes in 
Fig.~1.}
\end{figure}


\begin{thebibliography}{54}
\bibitem{ref36} It should be noted that the $X'$-, $X''$-, and $X'''$-integration
                involving $\lambda_\tau^{\rm dec}$ in (19) and (20) cannot 
                a priori be performed analytically, because of the appearance
                of the Earth's density $\rho(X,\theta)$.  It turns out, however,
                that the exact numerical integrations can be reproduced to 
                within 1\% by using the reasonable approximation \cite{ref4}
                $\rho(X,\theta)\simeq \rho_{av}(\theta)$ with the averaged
                $\rho_{av}(\theta)$ being not explicitly dependent on $X$
                (thus one can drop the explicit $X$ dependence in, for example,
                $\lambda_{\tau}^{\rm dec}(E,\, X',\, \theta')\simeq 
                \lambda_{\tau}^{\rm dec}(E,\, \theta)$ with $\theta'\equiv
                \theta(X')\simeq\theta(X)\equiv\theta$).  Notice that 
                $\rho_{av}(\theta)\equiv\int_0^{L}\rho\left(r(z,\theta)\right)
                \, dz/\int_0^L dz =X(\theta)/L$ with $r(z,\theta)=
                \sqrt{R_{\oplus}^2+z^2-zL}$ and $L=2R_{\oplus}\cos\theta$ 
                (see, e.g., \cite{ref15}).
                In particular the $X''$--integral in (20) receives its dominant
                contribution only from $X''$ being close to the upper limit of
                integration, $X''\simeq X'$.  Therefore the $X'''$--integral in
                (20) can be performed analytically.  It should be furthermore
                noted that, when (20) is inserted into (19), 
                $\lambda_{\tau}^{\rm dec}(E,X')$ in (19) gets cancelled to
                within a very good approximation (cf.~Eq.~(21)).
\end{thebibliography}
\end{document}